\documentstyle[epsfig,manuscript,prc,aps]{revtex}

\begin{document}

\title{Self-Consistent Hartree-Fock Based Random Phase Approximation And The 
Spurious State Mixing}

\author{B. K. Agrawal$^{1)}$, S. Shlomo$^{1)}$,  and A. I. Sanzhur$^{1,2)}$}
\address{$^{1)}$Cyclotron Institute, Texas A\&M University,
College Station, TX 77843-3366}
\address{$^{2)}$Institute for Nuclear Research, Kiev 03028, Ukraine}

\maketitle
\begin{abstract}
We use a fully self-consistent Hartree$-$Fock (HF) based continuum
random phase approximation (CRPA) to calculate strength functions $S(E)$
and transition densities $\rho_t(r)$ for isoscalar giant resonances with
multipolarities $L = 0$, 1 and 2 in $^{80}Zr$ nucleus.  In particular, we
consider the effects of spurious state mixing (SSM) in the isoscalar
giant dipole resonance (ISGDR) and extend  the projection method to
determine the mixing amplitude of spurious  state so that properly
normalized $S(E)$ and $\rho_t(r)$  having no contribution due to SSM
can be obtained.  For the calculation to be highly accurate we use a
very fine radial mesh (0.04 fm) and zero smearing width in HF$-$CRPA
calculations. We first  use  our most accurate results as a  basis to
establish the credibility of the projection method,  employed to eliminate the
SSM,  and then  to investigate the consequences of the common violation
of self-consistency, in actual implementation of HF based CRPA and discretized RPA (DRPA), as often
encountered in the published literature.
 The HF$-$DRPA calculations are carried out 
using a typical box size of 12 fm and a very large box   of 72
fm, for different values of
particle-hole  energy cutoff ranging from 50 to 600 MeV. 

\end{abstract}


\section{Introduction}

Hartree-Fock (HF) based random phase approximation (RPA) has been a very
successful theory in providing microscopic description of phenomena
associated with collective motion in nuclei \cite{Bohr75}. Accurate
information for important physical quantities can be obtained by comparing
the experimentally deduced strength function distribution, $S(E)$, with
the results obtained from HF-RPA theory. In particular, the strength
function distributions of the isoscalar giant monopole resonance (ISGMR)
and the isoscalar giant dipole resonance (ISGDR) are quite sensitive to
the value of the nuclear matter incompressibility coefficient, $K_{nm}$,
\cite{Bohr75,Stringari82,Shlomo93,Blaizot80}, a very important physical quantity
since it is directly related to the curvature of the equation of state.

Over the last two decades, a significant amount of experimental work
was carried out to identify  the strength distributions of the isoscalar
giant resonances in nuclei, particularly the ISGMR \cite{Shlomo93} and
ISGDR \cite{Clark01}. The main development in the area of experimental
investigation of the isoscalar giant resonances is the high accuracy
data, of excitation cross section, by $\alpha -$particle scattering,
obtained at Texas A\&M University using a beam analysis system
(BAS), a multipole-dipole-multipole (MDM) spectrometer and broad
range multiwire proportional counter. The new system improved the
signal to background ratio by more than a factor of 15. This led to
the discovery of a high lying structure in the strength function
of the ISGMR and the location of the ISGMR in light nuclei. Also,
accurate data on the ISGDR has been obtained for a wide range of
nuclei \cite{Clark01}. This has led to renewed  interest in the
nuclear response function and the need to carry out detailed and
accurate calculations of $S(E)$ and the transition density, $\rho_{t}$,
within the HF-RPA theory. In particular there have been quite a few
recent HF-RPA\cite{Hamamoto98,Kolomiets99,Gorelik00,Shlomo00,Colo00}
and relativistic mean field (RMF) based RPA\cite{Vret00,Piekarewicz00}
calculations of the ISGDR , considering the issues of (i) spurious state
mixing (SSM), (ii) the strength of the lower component (at 1$\hbar\omega$)
and (iii) the value of $K_{nm}$ deduced from the centroid energy $E_1$
of the ISGDR compression mode (at 3$\hbar\omega$).

Comparison between the recent data on the ISGMR and the results of
HF based RPA calculations confirms the value of $K_{nm} = 210 \pm 20$
MeV, determined earlier in \cite{Blaizot80}. It was first pointed out in
Ref. \cite{Dumitrescu83} that the HF-RPA results for $E_1$, obtained with
interactions adjusted to reproduce the ISGMR data, are higher than the
experimental values \cite{Morsch80,Djalali82} by more than 3~MeV and thus
this discrepancy between theory and experiment raises doubts concerning
the unambiguous extraction of $K_{nm}$ from energies of compression
modes. This discrepancy between theory and experiment was also reported
in more recent experiments \cite{Clark01,Davis97}. Recently, Shlomo
and Sanzhur\cite{Shlomo00} have addressed this discrepancy by carrying
out accurate microscopic calculations for $S(E)$ and the excitation
cross section $\sigma(E)$ of the ISGDR, within the folding model (FM)
distorted-wave-Born-approximation (DWBA), with $\rho_{t}$ obtained from
HF-RPA calculations and corrected for the SSM. They demonstrated that the
calculated $\sigma(E)$ drops below the experimental sensitivity in the
region of high excitation energy containing 30-40\% of the ISGDR energy
weighted sum rule (EWSR). This missing strength leads to a reduction of
more than 3.0 MeV in the value of $E_1$ and thus explains the discrepancy
between theory and experiment.

Clearly accurate calculations of $S(E)$ and $\sigma(E)$ are needed.
In fully self-consistent HF-RPA calculations, the spurious isoscalar
dipole ($T=0$, $L=1$) state (associated with the center of mass motion)
appears at energy  $E=0$ and no SSM in the ISGDR occurs.  It was pointed out in
\cite{Shlomo00} that none of the calculations carried out for $S(E)$ and
$\rho_t$ and published in the literature are fully self-consistent. In
some RPA calculations the mean field and the particle-hole interaction
$V_{ph}$ are chosen independently. Although this approach can provide
physical insight on the structure of giant resonances, it can not be used
to  accurately determine important physical quantity such as $K_{nm}$. In
self-consistent HF-RPA calculation\cite{Bertsch75} one starts by adopting
specific effective nucleon-nucleon interaction, $V_{12}$, such as the
Skyrme interaction, and carries out HF calculations. The parameters of
the interaction are determined by a fit to properties of nuclei (binding
energies, radii, etc.). Then one solves the RPA equation using the
particle-hole (p-h) interaction $V_{ph}$ which corresponds to $V_{12}$.
However, although not always stated in the literature, self-consistency
is violated in actual implementations of the RPA (and relativistic
RPA) calculations. One usually makes the following approximations:
(i) use a $V_{ph}$ which is not consistent with $V_{12}$. It is
common to neglect the two-body Coulomb and spin-orbit interactions
in $V_{ph}$ and approximate the momentum dependent parts in $V_{ph}$,
(ii) limiting the p-h space in a discretized calculation by a cut-off
energy $E_{ph}^{max}$, and (iii) introducing a smearing parameter (i.e.,
a Lorentzian with $\Gamma/2$). The consequences of these violations of
self-consistency on $S(E)$ and $\rho_t$ and of numerical inaccuracy are
usually ignored in the literature.

In this work we present results of detailed investigations of the
consequences of common violations of self-consistency in actual
implementations of HF based RPA, for determining the response
functions $S(E)$ and $\rho_{t}$ of isoscalar multipole (L=0,1 and 2)
giant resonances. In particular, we consider the ISGDR and concentrate
on the effects of the SSM.  We determine the effects of a violation
of self-consistency by comparing the calculated results for $S(E)$and
$\rho_{t}$ with those obtained from highly accurate fully self-consistent
HF- continuum RPA (HF$-$CRPA) calculations\cite{Shlomo75}. We also extend
the projection method for eliminating the effects of SSM, described
in Ref. \cite{Shlomo00}, to properly normalize $S(E)$ and $\rho_{t}$
and determine the mixing amplitude of the spurious state in the ISGDR.

In Section II we present an extension of the Green's function based
derivation of the projection operator method for eliminating the
effects of the SSM, described in \cite{Shlomo00}, to also account
for the proper normalization of the $S(E)$ and $\rho_t({\rm\bf r})$
of the ISGDR and determine the mixing amplitude of the spurious state,
obtained in HF-RPA calculations. We emphasize here that the method is
quite general and applicable for any scattering operator $F$ and for
any numerical method used in carrying out the RPA calculation, such as
configuration space RPA, coordinate space (continuum and discretized)
RPA and with and without the addition of smearing. We also provide in
this section the basic expressions used in the calculations and the
presentation of our results.

  In Section III we present and discuss
our results. We first present the results of a highly accurate  and
fully self-consistent HF$-$CRPA calculation of $S(E)$ and $\rho_t(r)$
in $^{80}Zr$, which we use as a basis for a comparison with results
obtained with common violations of self-consistency. These  accurate fully
self-consistent HF$-$CRPA results were obtained using $\Gamma = 0$ (i.e.,
no smearing) and very small mesh sizes of $dr_{\scriptstyle{HF}} = 0.04$
fm and $dr_{\scriptstyle{RPA}} = 0.04$ fm with corresponding number of
 mesh points $N_{\scriptstyle{HF}} = 900$ and $N_{\scriptstyle{RPA}} = 300$,
used in the HF and the CRPA calculations, respectively. We note that the
values of $S(E)$ and $\rho_t(r)$ associated with a bound RPA state were
deduced from the residue of the RPA Green's function.  Next, we present
our results of fully self-consistent HF$-$CRPA calculations (with $\Gamma
= 0$) carried out using various mesh sizes $dr_{\scriptstyle{HF}}$ and
$dr_{\scriptstyle{RPA}}$ and discuss the issue of  numerical accuracy. We
then present and discuss the results obtained with certain violations
of self-consistency in CRPA and discretized RPA (DRPA) calculations
and assess the effects on $S(E)$ and $\rho_t(E)$ by comparing with
the highly accurate fully self-consistent results over the whole
range of excitation energies. We point out that comparing the total
energy weighted transition strength with the EWSR may lead to incorrect
conclusions. 
Very recently the accuracy of the projection operator method in eliminating the effects of
the SSM on $S(E)$ and $\rho_t$   of the ISGDR was investigated in Refs.
\cite{Shlomo00,Hamamoto02}. However,  in these works, the calculations carried out using
mesh sizes  $dr \ge 0.1$ fm, were not fully self-consistent. We emphasize  that in the
present work we have carried out highly accurate  self-consistent  calculations,
established the accuracy of  the projection operator method and  provide assessments on the
effects  on $S(E)$, $E_L$ and $\rho_t$ of the isoscalar resonances with $L = 0$, 1 and 2,
which are due to  common violation of self-consistency  in actual implementation of
HF$-$RPA often encountered in the literature. We note that preliminary results of the
present work  were  presented earlier \cite{Agrawal02}.
In section IV we state our conclusion.

\section{Formalism}

The RPA Green's function $G$ \cite{Bertsch75,Shlomo75} is given by,  
\begin{equation}
G = G_0 (1+ V_{ph}G_0)^{-1}\ ,
\end{equation}
where $G_0$ is the free p-h Green's function given by,
\begin{equation}
G_0(\bbox{r},\bbox{r}',E)=-\sum_h \phi_h(\bbox{r})\left
[\frac{1}{H_0-\epsilon_h-\omega}+\frac{1}{H_0-\epsilon_h+\omega}\right
] \phi_h(\bbox{r}').
\label{equ:G0}
\end{equation}
Here $H_0$ is the HF hamiltonian and $\epsilon_h$ and $\psi_h$ are the
single particle energy and the wave function of the occupied state,
respectively. The continuum effects (particle escape width) are included
by using
\begin{equation}
G_{lj}(\bbox{r},\bbox{r}',E)=\frac{1}{H_0-E}=-\frac{2m}{\hbar^2}u_{lj}(r_<)
v_{lj}(r_>)/w,
\end{equation}
where $r_{<}$ and $r_{>}$ are the the lesser and the greater of $r$
and $r'$, respectively, $u$ and $v$ are the regular and irregular solution of
$H_{0}$, with the appropriate boundary conditions, respectively and $w$ is 
the Wronskian. The strength function $S(E)$ and transition density 
$\rho_t$, associated with the scattering operator,
\begin{equation}
F=\sum\limits_{i=1}^{A} f(\bbox{r}_i)\ ,
\label{equ:scop}
\end{equation}
are given by,
\begin{equation}
S(E)\!=\!\sum\limits_{n}\left|\langle 0 |F| n\rangle\right|^2
\delta(E-E_n)={1\over\pi}\mbox{Im}\left[\mbox{Tr}(fGf)\right],
\label{equ:strace}
\end{equation}
\begin{equation}
\rho_t(\bbox{r},E)={\Delta E\over\sqrt{S(E)\Delta E}}\int f(\bbox{r}\,')
\left[{1\over\pi}\mbox{Im}G(\bbox{r}\,',\bbox{r},E)\right]\,d\bbox{r}\,'\ .
\label{equ:rhot}
\end{equation}
Note that $\rho_t(\bbox{r},E)$, as defined in (\ref{equ:rhot}), is 
associated with the strength in the region of $E \pm \Delta {E}/2$ and
is consistent with 
\begin{equation}
S(E)=\left|\int\rho_t(\bbox{r},E)f(\bbox{r})\,d\bbox{r}\:\right|^2
\left/\Delta E\right.\ .
\label{equ:sfun}
\end{equation}
It is important to note that $S(E)$ and $\rho_t$ of a state at energy 
$E_n$ below the particle escape threshold (or having a very small width) 
can be obtained from Eqs. (\ref{equ:strace}) and (\ref{equ:rhot}), respectively, by 
replacing  ${1\over\pi}\mbox{Im}G(\bbox{r}\,',\bbox{r},E)$ with
\begin{equation}
\lim_{E \rightarrow E_n} \mbox{Re}G(\bbox{r}\,',\bbox{r},E)(E-E_n),
\label{equ:ReG}
\end{equation}

The energy weighted sum rule (EWSR) associated with the operator 
$f_{LM}=f(r)Y_{LM}$ is given by\cite{Bohr75},
\begin{equation}
EWSR(fY_{LM})= \int ES_{LM}(E)dE=
%\sum E_n \left|\langle 0 |F| n\rangle\right|^2\delta(E-E_n)=
{\hbar^2\over\ 2m}{A\over 4\pi}
\left[\langle 0|\left({df\over dr}\right)^2+
L(L+1)\left({f\over r}\right)^2|0\rangle\right].
\label{equ:ewsr}
\end{equation}
Using the equation of continuity and assuming that there is only one 
collective state \cite{Noble71,Deal73} with energy  $E_{coll}$, exhausting 100\% of the EWSR associated
with the scattering operator $f_{LM}=f(r)Y_{LM}$, one obtains the form
for the corresponding transition density,
\begin{equation}
\rho_t^{coll}(r)=-\frac{\hbar^2}{2m}\sqrt{\frac{2L+1}{EWSR(f_{LM})E_{coll}}}
\left[\left(\frac{1}{r}\frac{d^2}{dr^2}(rf) - 
\frac{L(L+1)}{r^2}f\right)\rho_0 +\frac{df}{dr}\frac{d\rho_0}{dr}\right].
\label{equ:collrho}
\end{equation}

Let us consider scattering operators, Eq.  (\ref{equ:scop}), with
\begin{equation}
f(\bbox{r})=f(r)Y_{1M}(\Omega)\ ,\ \ \ f_1(\bbox{r})=rY_{1M}(\Omega)\ ,
\end{equation}
and write ${\displaystyle{1\over\pi}}$Im$G$ as the sum of separable terms
\begin{equation}
R(\bbox{r}\,',\bbox{r},E)=
{1\over\pi}\mbox{Im}G(\bbox{r}\,',\bbox{r},E)=
\sum\limits_{n} d_{n}(E)\rho_n(\bbox{r})\rho_{n}(\bbox{r}\,')\ .
\label{equ:sepr}
\end{equation}
Note that $d_{n}(E)$ accounts for the energy dependence of 
$R(\bbox{r}\,',\bbox{r},E)$. In the case of a well defined resonance, or 
in a discretized continuum calculation, the sum in Eq. (\ref{equ:sepr}) 
has only one term. In this case $\rho_n$ is proportional to the transition 
density associated with the resonance and may contain a spurious state 
contribution. 
In general, due to the smearing with $\Gamma/2$, the sum 
in Eq. (\ref{equ:sepr}) may contain quite a few terms. We now write 
$\rho_n$ as
\begin{equation}
\rho_n(\bbox{r})=a_n\rho_{n3}(\bbox{r})+b_n\rho_{n1}(\bbox{r})\ ,
\label{equ:dssm}
\end{equation}
with
\begin{equation}
{a_n}^2 + {b_n}^2 = 1.0\, .
\end{equation}
Note that $\rho_{n1}(\bbox{r})$ is due to 
SSM and $\rho_{n3}$, associated with the ISGDR, fulfills the center of 
mass condition (for all $n$ )
\begin{equation}
\langle f_1\rho_{n3}\rangle=
\int f_1(\bbox{r})\rho_{n3}(\bbox{r})\,d\bbox{r}=0\ .
\label{equ:cond}
\end{equation} 
We point out that in the projection method for eliminating the effects of SSM, 
described in Ref. \cite{Shlomo00}, it was assumed that
 $a_n = 1.0$ (in (\ref{equ:dssm})). 

Following the derivation described in Ref. \cite{Shlomo00}, we first note
that all $\rho_{n1}$ coincide with the coherent spurious 
state transition density $\rho_{ss}(\bbox{r})$ \cite{Bertsch83}
\begin{equation}
\rho_{n1}(\bbox{r})=\rho_{ss}(\bbox{r})=
-\sqrt{\frac{\hbar^2}{2m}\frac{4\pi}{AE_{ss}}}
{\partial\rho_0\over\partial r}\,Y_{1M}(\Omega)\ ,
\label{equ:spur}
\end{equation}
where $E_{ss}$ is the spurious state energy and $\rho_0$ is the ground state 
density of the nucleus. Note that $\rho_{ss}$ in (\ref{equ:spur}) is
normalized to 100\% of the energy weighted sum rule (see (\ref{equ:ewsr})
and (\ref{equ:collrho})), 
\begin{equation}
EWSR(rY_{1M})=\frac{\hbar^2}{2m}\frac{3}{4\pi}A.
\label{equ:ewsr1}
\end{equation}
Looking for a projection operator that
projects out $\rho_{n1}(\bbox{r})$,
\begin{equation}
F_{\eta}=\sum\limits_{i=1}^{A} f_{\eta}(\bbox{r}_i) = F-\eta F_1\ ,
\end{equation}
with $f_{\eta} = f-\eta f_1$, we find that the value of 
$\eta$ associated with $\rho_{ss}$ is given by
\begin{equation}
\eta=\langle f\rho_{ss}\rangle/\langle f_1\rho_{ss}\rangle.
\label{equ:eta}
\end{equation}
Using (\ref{equ:cond}) and (\ref {equ:eta}) we have
\begin{equation}
S_{\eta}(E)=\langle f_{\eta}Rf_{\eta}\rangle=\langle fR_{33}f\rangle ,
\label{equ:seta}
\end{equation}
where,
\begin{equation}
R_{33}=\sum d_{n}(E)a_n^2\rho_{n3}(\bbox{r})\rho_{n3} 
(\bbox{r}\,') .
\label{equ:resp3}
\end{equation}

To determine $\rho_t$ for the ISGDR we first use (\ref{equ:rhot}),
(\ref{equ:sepr}), (\ref{equ:dssm}), (\ref{equ:cond}) and (\ref{equ:eta}) 
with $F_{\eta}$ and obtain
\begin{equation}
\rho_{\eta}(\bbox{r})={\Delta E\over\sqrt{S_{\eta}(E)\Delta E}}
\sum c_{n}a_{n}[a_n\rho_{a3}(\bbox{r})+
b_n\rho_{ss}(\bbox{r})]\ ,
\label{equ:tden}
\end{equation}
with $c_{n}=d_{n}(E)\langle f_{\eta}\rho_{{n}3}\rangle$.  
To project out the spurious
term from (\ref{equ:tden}) we make use of (\ref{equ:cond}) and obtain
\begin{equation}
\rho_t(\bbox{r})=\rho_{\eta}(\bbox{r})-b\rho_{ss}\ ,\ \ \ 
b=\langle f_1\rho_{\eta}\rangle/\langle f_1\rho_{ss}\rangle\ .
\label{equ:alph}
\end{equation}

To properly normalize $S_{\eta}(E)$ and $\rho_{t}$, we have to determine the 
mixing amplitudes $b_n$ of the spurious state in the ISGDR. These
amplitude can be obtained from the response function to the scattering
operator $f_1$. Using (\ref{equ:dssm}), (\ref{equ:cond}) and (\ref{equ:spur}) we obtained from (\ref{equ:sepr}),
\begin{equation}
S_{1}(E)=\langle f_{1}Rf_{1}\rangle=\langle f_1R_{11}f_1\rangle=
\sum d_{n}(E)b_n^2
\langle f_1\rho_{ss}\rangle^2.
\label{equ:resp1}
\end{equation}
Note that $\langle f_1\rho_{ss}\rangle$ can be obtained from the EWSR,
Eq. (\ref{equ:ewsr1}),
\begin{equation}
\langle f_1\rho_{ss}\rangle^2
={\hbar^2}{2m}\frac{3}{4\pi}A/E_{ss},
\label{equ:ewsr2}
\end{equation}
 and the SSM probabilities from 
\begin{equation}
b_n^2 = \frac{S_1(E_n)}{\langle f_1\rho_{ss}\rangle^2}\, .
\label{equ:bn}
\end{equation}

 In the present work we limit our discussion to the operator
$F_3=\sum\limits_{i=1}^{A} f_3(\bbox{r}_i)$, where 
$f(\bbox{r})=f_3(\bbox{r})=r^3Y_{1M}(\Omega)$. For this operator, the 
value of $\eta$ associated with the spurious transition density  
(\ref{equ:spur}) is
\begin{equation}
\eta={5\over 3}\langle r^2\rangle\ ,
\label{equ:eta3}
\end{equation}
 and 
\begin{equation}
S_\eta (E)= S_3(E) -2\eta S_{13}(E) + \eta^2 S_1(E),
\label{equ:seta1}
\end{equation}
where $S_3(E) = \langle f_3 R f_3\rangle$ is the strength function associated with $f_3$ and
$S_{13} = \langle f_1 R f_3\rangle$ is the non-diagonal strength function.

\section{Results and discussions}

In the following, we present our results for isoscalar giant resonances
 ($L = 0, 1$ and 2) obtained within the HF based RPA framework as briefly outlined in the
previous section. Calculations are performed for $^{80}Zr$ ($N = Z =
40$). The two-body
interaction $V_{12}$  is taken to be of a simplified Skyrme type,

\begin{equation}
\label{v12}
V_{12}=\delta(\vec{r}_1-\vec{r}_2)\left [
t_0+\frac{1}{6}t_3\rho^\alpha(\frac{\vec{r}_1+\vec{r}_2}{2})\right ],
\end{equation}
where $\alpha=1/3$, $t_0=-1800$ MeVfm$^3$ and  $t_3=12871$
MeVfm$^{3(\alpha+1)}$.  For these values of the interaction parameters
the nuclear matter  equation of state has a minimum at $E/A=-15.99$
MeV, $\rho_0=0.157$ fm$^{-3}$ with $K_{nm}= 226$ MeV, where $E/A$, $\rho_0$
and $K_{nm}$ being the binding energy per nucleon, matter  saturation density and
incompressibility coefficient for symmetric nuclear matter, respectively.
This choice of the two-body interaction enables us to use the continuum RPA method
to carry out a fully self-consistent calculation for giant resonances.
Following Ref. \cite{vau-prc5} one can write the mean field potential
$V_{mf}$ as,

\begin{equation}
\label{vmf}
V_{mf} = \frac{3}{4}t_0\rho(r) +\frac{\alpha+2}{16}t_3\rho^{\alpha+1}(r)
\end{equation}
and the particle-hole interaction $V_{ph}$ contributing to the isoscalar
channel is given by \cite{Bertsch75}
\begin{equation}
\label{vph}
V_{ph}=\delta(\vec{r}_1-\vec{r}_2)\left
[\frac{3}{4}t_0+\frac{(\alpha+1)(\alpha+2)}{16}t_3\rho^\alpha\right ].
\end{equation}

To begin with, we consider our results for  isoscalar giant monopole, 
dipole and quadrupole  resonances which are fully self-consistent and numerically
accurate. Then, we shall analyze the influence of various numerical
approximations on the centroid energies and transition densities for
these resonances.  Finally, we shall illustrate the possible effects of
the violation of self-consistency on the properties of these  isoscalar giant
resonances (ISGR).

\subsection{Self-consistent continuum RPA results}

We now present our results of fully self-consistent HF$-$CRPA calculations
for $^{80}Zr$, using the Skyrme interaction of Eq. (\ref{v12})
with spin-orbit and Coulomb interactions switched off.  It was
pointed out in \cite{Shlomo75} that in order to have cancellations
of the hole-hole transitions occurring in $G_0$ (Eq. (\ref{equ:G0}))
and obtain  numerically accurate results, it is important to employ
the same mean-field and the same integration algorithm for the bound
states and the single-particle Green's function, using a small mesh
size in double precision calculations.    In the following we first
present our results of highly accurate calculations obtained using
$dr_{\scriptstyle{HF}} = 0.04$ fm and $dr_{\scriptstyle{RPA}}=0.04$ fm,
and with no smearing ($\Gamma=0$ MeV), which we use in the following
as a basis for comparison with other calculations. We note that in
common implementations of HF-RPA one usually adopts the values of
$(dr_{\scriptstyle{HF}},dr_{\scriptstyle{RPA}})$=(0.1 fm, 0.3 fm) and
a  smearing parameter of  $\Gamma/2 \sim 1.0$ MeV.
 In the following we use the notation $dr =
(dr_{\scriptstyle{HF}},dr_{\scriptstyle{RPA}})$, with the values given
in units of fm.

To facilitate our discussions we have displayed in Table \ref{sp-energy}
the HF single-particle energies for $^{80}Zr$ obtained for
$dr_{\scriptstyle{HF}} = 0.04$ fm. In Table \ref{exact-ewsr} we give the
values  for the density radial moments  $\langle r^2\rangle $, $\langle
r^4\rangle$ and  EWSRs (Eq. (\ref{equ:ewsr})) for various multipoles
evaluated for different values of mesh size in the HF calculation.  In
Table \ref{str-0404} we present the values of energy weighted transition
strengths (EWTS) for free and CRPA responses  obtained using the operators
$f_3$, $f_1$ and $f_\eta$ with $dr =( 0.04,0.04)$  and $\Gamma = 0$
MeV. The quantities $S_1^{\scriptstyle{EW}} $, $S_3^{\scriptstyle{EW}}
$, $S_{13}^{\scriptstyle{EW}} $ and $S_{\eta}^{\scriptstyle{EW}} $ in
Table \ref{str-0404} denote the EWTS  for the corresponding strength
functions $S_1$, $S_3$, $S_{13}$ and $S_\eta$, respectively,  see
Eq. (\ref{equ:seta1}).  The transition strengths associated with sharp
transitions were determined from the residues of the Green's function,
using its real part (see Eq. (\ref{equ:ReG})).  For the free response
we get sharp peaks at the bound state  single particle-hole transitions
associated with $L = 1$.  These transitions can be easily identified from
Table \ref{sp-energy} as $0g\rightarrow 0f$  (10.83), $1d\rightarrow 1p$
(11.35), $2s\rightarrow 1p$  (12.70), $1d\rightarrow 0f$  (17.43),
$1d\rightarrow0p$  (35.16), and $2s\rightarrow 0p$  (36.52), with
corresponding transition energies given in brackets in MeV.  We checked
that  the values of the EWTS  for these sharp transitions agree with
the  corresponding values obtained directly from the particle and hole
wave functions.

For  CRPA response, the sharp peaks occur below the particle threshold at
15.33  MeV.  In addition to these sharp transitions, we have contributions
from the continuum starting at the particle threshold.  We obtained the
contributions from the continuum by integrating the energy weighted
strength function using small enough energy steps of $dE = 0.01$ MeV.
It is seen  from Table \ref{str-0404} that the spurious state mixing is
significantly larger for the free response (see 3rd column). Once the
spurious state mixing is eliminated using the projection operator $f_\eta$
we find from the 2nd and 5th columns of this table that most of the
strengths of the free response in the $1\hbar\omega$ region of excitation
energy  ($E <  20$ MeV) is   spurious in nature. Only $6.8\%$ of the EWTS
for the operator $f_3$ contributes to the intrinsic excitations for $E <
20$ MeV. On the other hand, in  the case of CRPA, since the calculation
is fully self-consistent and numerically very accurate, the resonance
occurring at 0.079 MeV is fully spurious and it exhausts $99.99\%$
of the  EWSR  associated with the operator $f_1$.  For $E > 0.08$ MeV,
the values of $S_1$ and $S_{13}$ are so small that SSM is negligible. We
see from the Table \ref{str-0404} that the  values of $b_n^2$  is $\sim
10^{-8}$ (see Eq.( \ref{equ:bn})) indicating that the  SSM is so small
that one need not renormalize the strength $S_\eta$.  For $E > 0.08$
MeV, the values of the CRPA EWTS for the operators $f_3$ and $f_\eta$
are the same within $1\%$. We would like to emphasize that though the
spurious state mixing is significantly large for the free response, it is
fully eliminated by using the projection operator $f_\eta$ giving rise to
$99.95\%$ of the expected EWSR which is quite close to the CRPA results.
It may also be added that the fraction energy weighted sum rule, FEWSR =
EWTS/EWSR,   for the operator $f_\eta$ is $8.4\%$ and $7.4\%$ for $E <
20$ MeV in the case of free and CRPA responses, respectively.

A proper test of  a fully self-consistent calculation is to check how
close  $\rho_t(r,E_{ss})$ is to $\rho_{ss}$, where $\rho_t (r,E_{ss})$
is obtained from Eqs. (\ref{equ:rhot}) and (\ref{equ:ReG}) at  the
spurious state  energy $E_{ss}$   using $f_1$. In Fig.  \ref{rho-ss}
we compare the CRPA result for the $\rho_t(r,E_{ss})$  with the coherent
state transition density calculated using Eq. (\ref{equ:spur}).
It is seen in Fig. \ref{rho-ss} that in this highly accurate HF$-$CRPA
calculation $\rho_t(r,E_{ss})$ coincides with $\rho_{ss}$, indicating a
very negligible SSM.

We shall now present some plots for the strength functions for various
multipoles obtained from our most accurate calculations. For plotting
purpose we used a very small smearing  width $\Gamma /2= 0.025$ MeV. In
Figs.  \ref{isgdr-free} and \ref{isgdr-rpa} we have shown the free and RPA
response for the ISGDR, respectively.  We see from  Fig. \ref{isgdr-free}
that most of the spurious components lie in the low energy region ($E <
20$ MeV). As mentioned before, we see that  the response for the operators
$r^3$ and $(r^3-\eta r)$ are indistinguishable in the  case of a fully
self-consistent HF based CRPA calculation.  It  also appears from these
figures that particle-hole correlations  do not alter the ISGDR strength
distribution (shown in Fig. \ref{isgdr-free}) very much which suggests
that the isoscalar  dipole state is not a  very collective one. In
Figs. \ref{isgmr} and \ref{isgqr} we have shown the plots for the ISGMR
and ISGQR response functions, respectively. We have also carried out
these calculations for $dr=(0.24,0.24)$ (not shown here)  and find that
they can not be distinguished from our most accurate calculations. We
also note that at the surface the transition density for ISGMR looks like
$3\rho_0+rd\rho_0/dr$ as given by Eq.  (\ref{equ:collrho}), whereas,
the   ISGQR transition density looks more like $d\rho_0/dr$ rather
than $rd\rho_0/dr$ as given by Eq. (\ref{equ:collrho}).  We point out
that Eq. (\ref{equ:collrho})  was derived under the assumption that one
collective state exhausts the EWSR.

We have repeated the fully self-consistent  calculations for $\Gamma
= 0$ MeV, using various values  of $dr_{\scriptstyle{HF}}$ and
$dr_{\scriptstyle{RPA}}$. In Table \ref{str-2424-0424} we present
CRPA results for the EWTS  only for $dr= (0.24,0.24)$ and $ (0.04,0.24)$ with
$N_{\scriptstyle{RPA}}=50$.  We see from Table \ref{str-2424-0424} that
the results for the  operator $f_3$ for the different combinations
of the mesh size differ by about $2.5\%$.  The spurious state for
$dr=(0.24,0.24)$ occurs at 0.7 MeV and its excitation energy  becomes
imaginary for $dr =( 0.04,0.24)$. By multiplying the particle-hole
interaction by a constant factor $V_{sc} = 0.9916$ we push the spurious
state to 0.1 MeV for $dr = (0.04,0.24)$ calculations.  Nevertheless,
we find that the SSM, or equivalently  $b_n^2$ is very small ($\sim
10^{-6}$).  Once the spurious components are eliminated using the
projection operator $f_\eta$, we get  $99.40\%$ and $99.76\%$ of the
expected EWSR for $dr = (0.24,0.24)$ and $(0.04,0.24)$, respectively. So
far we have demonstrated that (i) as long as the calculation is fully
self-consistent and numerically highly accurate, there is practically
no spurious state mixing and (ii) the spurious state mixing introduced
due to the use of a large mesh size  (0.24 fm) in a CRPA calculation
can be projected out using the operator $f_\eta$.

 In Tables \ref{centroids} and \ref{ewsr} we have collected
the centroid energies and the FEWSR, respectively,  for the isoscalar
resonances with $L = 0$, 1 and 2 calculated using different combinations
of the mesh size and a fixed value of $\Gamma/2 = 0.025$ MeV. We  notice
that as long as the particle-hole interaction is not renormalized (i.e.,
$V_{sc}=1.0$) the centroid energies of the resonances do  not deviate by
more than 0.5$\%$ compared with the most accurate values.  Though  the
energy of spurious state  is sensitive to the values of the mesh sizes
and increases  from $0.08$ MeV to 0.71 MeV with the increase of radial
mesh size from  0.04 fm to 0.24 fm,  the centroid energy  for ISGDR
changes  only by about 0.08 MeV.  Even if $V_{sc}$ is used to shift
the spurious peak to 2.0 MeV, the centroid energy for $L=0$ and $L=1$
resonance do not change appreciably. However, the centroid energy for
the  $L=2$ resonance goes up by about $2\%$ (0.3 MeV) . It is also clear
from Table \ref{centroids} and Figs. \ref{isgdr-rpa}, \ref{isgmr} and
\ref{isgqr} that  the peak energy for ISGMR and ISGDR is higher
than their centroid energies by about 0.5 and 0.15 MeV, respectively.
From the Table \ref{ewsr} we find that when $dr =  (0.04,0.24)$, the
values of the total EWTS for ISGMR and ISGQR are overestimated by $1 - 2\%$.

\subsection{Influence of the smearing parameter $\Gamma$}

One of the requirements  to avoid any SSM is that one must not use any
smearing parameter (i.e., $\Gamma=0$) and the calculations should be
performed using a very fine mesh in the co-ordinate space while solving
HF and RPA equations. However, one typically uses $\Gamma/2 \sim 1.0$
MeV and the  mesh $dr \ge 0.1$  fm.  If the smearing width is finite,
the spurious state would have a long energy tail which can give rise to
large SSM. Because, $\rho_{ss}\propto d\rho_0/dr$,  which is a  surface
peaked functions,  and has a  large matrix element for the operator
$f_3$. One must project out the SSM by making use of the projection
operator $f_\eta$.

In Fig. \ref{isgdr-gam2} we plot CRPA results  for the spurious state
and ISGDR strength functions  calculated using radial mesh size of
0.04 fm and smearing parameter $\Gamma/2 = 1$ MeV. We clearly see
from the figure that the strength function for the  spurious state is
extended up to a very high energy. The SSM caused due to the energy
tail of the  spurious state is eliminated using the operator $f_\eta$.
In Table \ref{ewsr-isgdr} we give  the values of FEWSR,  associated
with the scattering operator $f_\eta$,  for the ISGDR for various
energy ranges up to 150 MeV obtained using different values for the
mesh size and the smearing parameter in the HF$-$CRPA calculation.
Considering the values of the FEWSR in each energy range $\omega_1 -
\omega_2$ of Table \ref{ewsr-isgdr} it can be easily seen that these
values are practically the same as those obtained with $\Gamma = 0$,
i.e., the SSM due to non-zero $\Gamma$ is completely projected out.
For $\Gamma/2 = 1.0$ MeV the values for FEWSR for  $E = 0 - 18$ MeV
is lower by about $1\%$ as compared to that for $\Gamma = 0$. This is
because of the resonance at $\sim 17.0$ MeV (see Fig. \ref{isgdr-rpa}).
If we integrate the energy weighted strength for $E = 0 - 20$ MeV, this
difference reduces from $1\%$ to about $0.5\%$. We also note that for
$\Gamma/2 = 1.0$ MeV the total FEWSR obtained by integrating up to $E =
150$ MeV is about $1\%$ lower than the one obtained  for $\Gamma = 0$. Of
course, this is because of the remaining strength beyond 150 MeV. For
instance, in the  case of $dr =(0.24, 0.24)$  and $\Gamma/2 = 1.0 $
MeV we get FEWSR = 0.48$\%$  for  the region for $E = 150 - 300$ MeV.

We point out  that due to $\Gamma \not= 0$, the transition density  $\rho_t$  
 calculated using Eq. (\ref{equ:rhot})  depends on the scattering
operator  $f$. The consequences of this on the  $S(E)$  and $\rho_t$ of the ISGDR
was investigated and discussed in
detail in Ref. \cite{Shlomo00} and we do not repeat it here.  We thus demonstrated  that using the
projection scattering operator $f_\eta$ one can accurately eliminate
the SSM effects occurring due to the use of a finite  smearing parameter
$\Gamma/2$.

\subsection {HF$-$DRPA results}

 The continuum can be discretized by confining the nucleus in a box
of finite size. One can satisfactorily reproduce the continuum results,
provided the calculations are carried out  using a box of very large size
(i.e., dense discretization) and the cut-off for the  particle-hole
excitation energy ($E_{ph}^{max}$) set to be reasonably high.
We now consider our results obtained  by discretizing the continuum
using boxes of different sizes. The length of the box is given by
$N_{\scriptstyle{HF}}$ times $dr_{\scriptstyle{HF}}$, where  $N_{HF}$
is the number of radial mesh point used in  a HF calculation. In the
following, we present the results for  discretized RPA calculations
obtained using $dr = (0.08,0.24)$ with $N_{\scriptstyle{HF}}  =
150$ and 900 (box sizes of 12 and 72 fm, respectively). In Figs.
\ref{isgdr-drpa}a, and \ref{isgdr-drpa}b  we show  the ISGDR response
for box sizes of 12 and 72 fm and smearing parameter $\Gamma/2 =
0.25$ and 1.0 MeV, keeping $E_{ph}^{max}=200$ MeV, together with
the corresponding results obtained in HF$-$CRPA.  We see that the
DRPA results obtained for the larger box coincide with the results
obtained  within CRPA.  The transition  strength gets fragmented if the
discretization  is carried out using a small box. To avoid  misleading
interpretation of the fragmentation and obtain agreement with the CRPA
results,  one needs to use a  larger value of the smearing parameter,
consistent with the size of  the box.  To examine more closely the
effects of discretization on the response function we  present  in
Table \ref{ewsr-drpa} our DRPA results for  the FEWSR over various
energy ranges up to 150 MeV. It is evident from this table that the
total FEWSR increases significantly  when $E_{ph}^{max}$ is increased
from 50 MeV to 200 MeV. This increase is about $5 - 6\%$ and  $9 - 10\%$
for $\Gamma = 0.5$ and 2 MeV, respectively.  With a further increase in
$E_{ph}^{max}$ there is no noticeable change in the value of the total FEWSR. It
can be easily verified from this table that the FEWSR associated with
the low-lying ISGDR component($E  <  20 $ MeV) increases from  $6.4\%$
to $6.9\%$ when $E_{ph}^{max}$ is increased from 50 to 200 MeV for
the case of $N_{\scriptstyle{HF}}=150$ and $\Gamma = 0.5$ MeV and it
further increases to $7.2\%$ for $N_{\scriptstyle{HF}}=900$ (see also
Table \ref{ewsr-isgdr}).  Comparing the Tables \ref{ewsr-isgdr} and
\ref{ewsr-drpa} we can conclude (see also Figs.  \ref{isgdr-drpa}a and
\ref{isgdr-drpa}b) that with the proper choice of discretization  and
$E_{ph}^{max}$ one can mimic the continuum even for smaller values of
$\Gamma \sim 0.5 - 1.0$ MeV.  Comparing the values of FEWSR,  in each
of the energy range $\omega_1 - \omega_2$,  of Table \ref{ewsr-drpa}
with Table \ref{ewsr-isgdr} we conclude that using $f_\eta$ one
accurately eliminates SSM  occurring due to the  use of a low value
for $E_{ph}^{max}$.

In Table \ref{ecen-eph} we have  displayed the values of $E_{ss}$
and the centroid energies for the $L = 0$, 1 and 2 isoscalar giant
resonances. These results are obtained using $N_{\scriptstyle{HF}} =
900$, $\Gamma/2 = 0.25$ MeV with different values of $E_{ph}^{max}$ in
HF-DRPA calculations.  The corresponding HF$-$CRPA results are given in the last row of the table.
We clearly see that as $E_{ph}^{max}$ increases,
the centroid energies  $E_0$, $E_1$ and $E_2$ converge to their corresponding exact values
obtained using HF-CRPA.  
However,  this convergence is slower for the spurious state energy $E_{ss}$.
For low values of $E_{ph}^{max}$ we observe
that the centroid energy  for ISGMR is overestimated by  0.5 MeV, which
can significantly effect the value of nuclear incompressibility. We also
notice that $E_1 =  35.3$ MeV is little low for $E_{ph}^{max} = 50$ MeV,
because of the fact that the resonance energy for the ISGDR  compressional mode
is about 35.5 MeV (see also Fig. \ref{isgdr-rpa}).

We saw in the previous subsection that the spurious transition
density $\rho_t(r,E_{ss})$ obtained using a fully self-consistent CRPA
calculation is indistinguishable from the corresponding collective
model form for  $\rho_{ss}$ which is proportional to $d\rho_0/dr$. In
Fig. \ref{rhoss-drpa} we show some of the DRPA results for $\rho_{t}
(r,E_{ss})$ and compare them with the $\rho_{ss}$. We see that for
$E_{ph}^{max} = 50$ MeV $\rho_t(r,E_{ss})$ deviates from $d\rho_0/dr$
even for the case of $N_{HF} = 900$.  However,  for $E_{ph}^{max} = 200$
MeV, the   $\rho_t(r,E_{ss})$  from  the DRPA is almost identical to the
collective model results. Thus, one must use
 a reasonably large value
for the cut-off energy, $E_{ph}^{max}$, in order to fully eliminate from
the intrinsic excitations the contribution due to SSM.

\subsection{Effects of violation of  self-consistency}

\label{selfc-violation}

So far we have examined the various effects of numerical   approximation
on the properties of  the isoscalar giant resonances of multipolarity 
$L = 0 - 2$ and established the validity of the projection operator
method in eliminating the SSM effects from the ISGDR.  Here we
report our investigations of the influence of  certain violations of
self-consistency on the strength function for isoscalar giant monopole
($L=0$), dipole ($L=1$) and quadrupole ($L=2$) resonances. These
investigations are quite important in view of the fact that one often
performs  non self-consistent calculations for giant resonances such as
the  use of a  phenomenological nuclear mean field (e.g., Woods-Saxon
potential) and Landau-Migdal particle-hole interaction\cite{Gorelik00}.
Moreover, one often come across HF$-$RPA calculations  carried  out using
particle-hole interaction not consistent with the mean field used in HF.
We present below the results for HF based CRPA calculations carried
out with  the two-body interaction given in Eq. (\ref{v12}).   We use
the  parameter $V_{sc}$ to renormalize the particle-hole interaction
(i.e., $t_0\rightarrow t_0V_{sc} $ and $t_3\rightarrow t_3V_{sc}$ in
Eq. (\ref{vph})) so that the  position of the  spurious state can be
adjusted close to zero. To study the consequences of the violation of
self-consistency we vary $t_0$  and $t_3$ only in the particle-hole
interaction (only in Eq. (\ref{vph})).  

In Table \ref{vzz-vs-ecen}
we summarize our results for the centroid energies for isoscalar giant
resonances for $L = 0 - 2$.  The quantity $K_{nm}^\prime$ is the nuclear
matter incompressibility coefficient associated with the  renormalized
parameters  $t_0V_{sc}$ and $t_3V_{sc}$. Here, $t_0$ and $t_3$ are the
values used in Eq. (\ref{vph}).  Let us first  consider the results
obtained by varying $t_0$ by $\pm 5\%$ and $\pm 10\%$ and keeping $t_3
= 12871$ MeVfm$^4$.  It can be clearly seen from the table that the
centroid energies for ISGMR and ISGDR significantly differ from their
corresponding self-consistent values even if $V_{sc}$ is adjusted to give
$E_{ss} = 0.1$ MeV.  On the other hand, the centroid energy for ISGQR
reattains its self-consistent value when $V_{sc}$ is adjusted to yield
$E_{ss} = 0.1$ MeV.  One may understand this discrepancy in terms of the
incompressibility coefficient. With the renormalization of $V_{ph}$,
though, $E_{ss}$ becomes close to zero, but values of $K^\prime_{nm}$
in the RPA calculation  remain quite different then the HF value of
226 MeV.  In Fig. \ref{e0-e1-kph} we plot the values of $E_0$ and $E_1$
versus $\sqrt{K^\prime_{nm}}$ for the cases with $E_{ss} = 0.1$ MeV.
This plot clearly depicts the systematic increase in $E_0$ and $E_1$
with increase in $K^\prime_{nm}$.  One  may be  tempted  to infer at this
point that as long as the nuclear matter incompressibility associated
with the particle-hole interaction and the mean field is the same,
centroid energies for  the resonances considered here  may come out to
be reliable. In  order to verify this, we adjust $t_3$  in particle-hole
interaction in such a way that $K_{nm}^\prime$ becomes 226 MeV when $t_0$
is varied by $\pm 10\%$. We see from Table \ref{vzz-vs-ecen} that  even
if $K_{nm}^\prime$ is adjusted to   226   MeV, the values of $E_0$ and
$E_1$  are  off by about $10\%$ and $3.5\%$, respectively. This is due
to the fact that the shape of the particle-hole interaction  is not the
same, though,  the $K_{nm}^\prime $ is
kept constant.
We note that if the ISGMR  centroid energy is determined within
$10\%$ accuracy, the value of nuclear
matter incompressibility will be correct only within $20\%$.

Apart from the centroid  energies for the giant resonances, it is also
important to investigate  the effects  on the strength function itself
when the self-consistency is not maintained.  We looked into the plots
for the strength  functions $S$ and $S_\eta$ for the operators $f_3$
and $f_\eta$, respectively, for the different cases listed in Table
\ref{vzz-vs-ecen}.  We find that $S_3 \ge  S_\eta$ or  $S_3 < S_\eta$,
depending on the sign of interference between the spurious state  and
the intrinsic state (i.e., sign of the non-diagonal strength $S_{13}$). As
an illustrative example, we show in Fig. \ref{isgdr-tsc}  our results
for the case in which $t_0$ is varied by $-10\%$ and $V_{sc} = 1.7118$.
Similar is the case when $t_0$ is varied by $-5\%$ and $V_{sc} =
1.2938$. These values of $V_{sc}$  were chosen so that $E_{ss} = 0.1$
MeV.  In Figs. \ref{isgr-selfcv}a, \ref{isgr-selfcv}b and  \ref{isgr-selfcv}c  we compare the
fully self-consistent results for isoscalar giant resonances with those
obtained by varying $t_0$ by $\pm5\%$ in Eq. (\ref{vph}) and $V_{sc}$
is adjusted to yield $E_{ss} = 0.1$ MeV. We see that the strength
function for ISGMR and ISGDR are significantly different compared with
their corresponding self-consistent results.  Whereas, in case of ISGQR
not only their centroid energies, but also the strength function seem
to agree well with the corresponding  self-consistent results.  It is
very important to point out that the violation of self-consistency causes
redistribution of the strength  in such a way that the total EWTS remains
unaltered. This redistribution may be crucial in determining the energy
weighted strengths associated with the low energy  and the high-lying energy  components of
the ISGDR. For example, the fraction of the EWSR (in percent) for the
energy range $E = 0 - 20 $ MeV  is 6.94, 9.33 and 12.42 for  $t_0 =
-1710$, $-1800$ and $-1890$ MeV fm$^{3}$, respectively, and for $E =
0 - 150$ MeV we have for the FEWSR = $99.76\%$ in these three cases.

We now focus on the influence of self-consistency violation when the
continuum is discretized. As  seen above, the discretization introduces
two additional constraints, namely, the box size 
used in HF calculations
and the maximum  allowed particle-hole energy $E_{ph}^{max}$. We
present here only the results for box size of 12 fm  with
$E_{ph}^{max} = 50 $ and $200$ MeV. In Fig. \ref{isgdr-selfcv-drpa1}
we compare the ISGDR response function  obtained for $t_0 = -1620$,
-1800 and -1980 MeVfm$^3$,  keeping $E_{ph}^{max} = 50$ MeV. 
Similar results are shown in Fig. \ref{isgdr-selfcv-drpa2} but  obtained by 
raising $E_{ph}^{max}$ to 200 MeV. From $S_3(E)$ we see clearly that when the
particle-hole interaction is in accordance with the mean-field potential,
the SSM is only due to $\Gamma\not=0$. For the cases with $t_0 \not=
-1800$ MeV fm$^3$ one can immediately  see a marked enhancement  in
spuriocity at $E = 10 - 12$ MeV. Furthermore, it is startling to see
that the total FEWSR associated with operator $f_\eta$ for $t_0=-1620$
and $-1980$ MeVfm$^3$ is 94.97$\%$ and $58.97\%$, respectively, and it
is 95.13 $\%$ for $t_0 = -1800$  MeVfm$^3$.  For $t_0=-1710$  MeVfm$^3$
we get a total FEWSR = $88.39\%$ which is once again too much  off
compared to its expected value. We repeated the same analysis for
box size 72 fm  keeping $E_{ph}^{max} = 50$ MeV but did not
find any appreciable change in the values of the  total FEWSR. However,
when we raised the $E_{ph}^{max}$ from 50 to 200 MeV, we get the total
FEWSR $99.63\%$, $100.52\%$ and $99.94\%$ for $t_0 = -1620$, $-1800$ and
$-1980$ MeVfm$^3$, respectively.  

We also calculate  the  SSM probabilities  (i.e.,
$b_n^2$)  when self-consistency is not maintained. The values of $b_n^2$
are extracted using an  extremely small  smearing parameter.  In case of
$t_0 = -1620$ MeVfm$^3$  and $E_{ph}^{max} = 50$ MeV  used in DRPA calculation,
we find that $E_{ss} = 9.84$ MeV. We get from Eq. (\ref{equ:bn}),
 $b_n^2 = 2.4\%$
for the state  occurring at  $\sim 13$ MeV. When $V_{sc}$
is  adjusted to push  the spurious state  energy  $E_{ss}$ to about $0.1$ MeV, the EWTS
 of the 13 MeV state, associated with SSM, remains  unchanged. Consequently  $b_n^2$ reduces by
two orders of magnitude. We  thus conclude that since the values of $b_n^2$ are less than
a few percent even with large violation of self-consistency, the renormalization of the
strength function is not needed.

We  have considered the effects  on   the ISGDR strength function when
Coulomb/spin-orbit interaction is switched on in the HF calculation,
but, ignoring it in the particle-hole interaction.  We find that when
spin-orbit interaction is included, the strength function obtained
using $\Gamma/2 = 1$ MeV  hardly gets affected at any energy and  the differences can
not be seen on the plots (not shown here).  This is due to the fact
that the  nucleus in question, $^{80}Zr$,  is spin saturated, i.e.,
the single-particle states with $j = l\pm 1/2$ are occupied. However,
this may not be the case for non spin-saturated heavy nuclei.
 When we
carried out similar exercise with the Coulomb interaction, the mean
field changes significantly  and   we find that the strength functions
gets  shifted towards  lower energy by about 2.0 MeV.  We note that,  with the
inclusion of Coulomb interaction, the particle threshold for protons
reduces from 15.33 MeV to 3.5  MeV.

\section{Conclusions}

We have carried out  self-consistent HF based CRPA calculations for
isoscalar giant resonances   with multipolarities $L = 0$, 1 and 2 for
$^{80}Zr$ nucleus as an example. We demonstrate that if a self-consistent
calculation is performed using zero smearing width and a very fine radial
mesh size ($dr = 0.04$ fm), the spurious state occurs at $E_{ss} = 0.08$
MeV and  the ISGDR response  for the operators $f_3$ and $f_\eta $ are
essentially the same for the energy $E > E_{ss}$ which indicates  no
SSM and the corresponding EWSR is reproduced remarkably well.  When  we
use $dr = 0.24$ fm in HF and CRPA calculations, $E_{ss}$ becomes about
0.7 MeV and there exists a small SSM. 
The amplitude of this SSM (i.e., $b_n^2$) is  $\sim 10^{-6}$,  which is   negligibly small
and one need not renormalize the projected strength function.
  Although  the position of the
spurious state is quite sensitive to the radial mesh size and smearing
parameter $\Gamma$,  the centroid energy for the isoscalar resonances for $L =
0$, 1 and 2 do not change by more than $0.5\%$.

We have also performed the calculation for $L = 0$, 1 and 2 isoscalar
resonances by discretizing the continuum using boxes of different sizes
(12 and 72 fm) with $E_{ph}^{max}$ ranging from $50 - 600$ MeV. We
found that the strength distribution  is fragmented over a wide energy
range for the case of the  smaller box irrespective of $E_{ph}^{max}$.
For the case of discretization in a large box (72 fm) with $E_{ph}^{max}
= 200$ MeV we find that the strength distribution agrees  reasonably well
with the corresponding   one obtained from CRPA, if a moderate value of
the smearing parameter ($\Gamma /2 \sim 1$ MeV) is used.  The spurious
state occurs  at about $4.5$ MeV  for $E_{ph}^{max} = 50$ MeV for both
the small as well as  large box  discretization considered. With the
increase of $E_{ph}^{max} $ to 600 MeV, we find that $E_{ss}$ approaches
the corresponding value obtained within the  CRPA. The centroid energies
for $L = 0$, 1 and 2 resonances  converge  to their corresponding exact
values obtained from HF$-$CRPA. This convergence is little slow in case
of spurious state energy. For $E_{ph}^{max} = 50$ MeV, the transition
density  $\rho_t(r,E_{ss})$ at the spurious state energy obtained
using discretized RPA  differs from  the corresponding CRPA results
(which reproduce $\rho_{ss}$). However, with increase of $E_{ph}^{max}
$ to   200  MeV,  DRPA  results for the spurious state transition density
$\rho_t(r,E_{ss})$ become  quite
 close to the CRPA results.  We also point out  that
one should use  $E_{ph}^{max} \ge 200$ MeV in order to calculate  the
centroid  energies of the isoscalar $L = 0$, 1 and 2 resonances with
the accuracy of 0.1 MeV, comparable  to the experimental   uncertainties.

We have demonstrated that  the spurious state mixing  due to the
non-zero smearing width and a choice of a coarse sized radial mesh
can be accurately eliminated using the projection   operator $f_\eta$.
Furthermore, we show that the SSM due to  a  small value of $E_{ph}^{max}$
used in the DRPA calculation can be fully eliminated by applying the
projection method.

We have investigated  the consequences of violation of self-consistency
on the $S(E)$ and $\rho_t$ of the isoscalar $L = 0$,1 and 2 giant
resonances by varying the parameter $t_0$ by $\pm 5\%$ and $\pm 10\%$
in the patrticle-hole interaction. We find that if the self-consistency
is not maintained then the  values of $E_{ss}$ and centroid energies
for the $L = 0$ , 1 and 2 isoscalar resonances  are significantly
different compared with their self-consistent values.  Even if the
particle-hole interaction is renormalized to shift $E_{ss}$ close
to its self-consistent value, the centroid energies for $L = 0$ and
1 resonances could not  be corrected. This is due to the fact that
though the renormalization corrects the value of $E_{ss}$, the nuclear
matter incompressibility coefficient, $K_{nm}^\prime$  associated with
the particle-hole interaction is quite different than the one with the
mean-field.  However, the $L = 2$ resonance is not very  sensitive to  the
self-consistency violation as long as the particle-hole interaction is
renormalized to shift $E_{ss}$ close to its self-consistent value.  It is
also important to point out that the violation of self-consistency  causes
a significant  redistribution  of the transition strength. In particular,
energy weighted transition strength of the lower energy component  ($E
< 20$ MeV)  of the   ISGDR response function may differ  by $50\%$.
The values of the SSM  probabilities  $b_n^2$  were found to be less
than  $1 - 2\%$. Therefore, one can  neglect the renormalization of the
ISGDR strength function obtained using projection operator $f_\eta$.
Further, we found that the total energy weighted transition strength
for the operator $f_\eta $ remains unaltered even with the violation
of self-consistency.

Calculations were also carried out by changing the parameters appearing
in the particle-hole interaction in such a way that the nuclear matter
incompressibility coefficient  associated with it remains  unaltered.  We find that
though the incompressibility coefficient  associated with  the  particle-hole and
the mean field are kept the  same, due to the lack of self-consistency,
the centroid energy for $L = 0$ and 1 isoscalar resonances  are off by
$10\%$  and  $3.5\%$, respectively, compared to their self-consistent values.
We may remark  that  if the ISGMR centroid energy is determined  with accuracy of 
$10\%$, the value of $K_{nm}$   deduced from a comparison  with experimental data is  then accurate  within only $20\%$.

This work was supported in part by the US Department of Energy under 
grant no. DOE-FG03-93ER40773.

\begin{figure}
\caption{\label{rho-ss} Comparison  of fully self-consistent HF$-$CRPA result
for spurious state transition density (in arbitrary units)   obtained using 
operator $f_1$ in Eq. (\ref{equ:rhot}) with the corresponding coherent
state transition density .
The HF$-$CRPA calculation is carried using radial mesh size  $dr = (0.04, 0.04)$  with
no smearing width ($\Gamma = 0$  MeV).} 

\vspace*{0.25 true in}
\caption{\label{isgdr-free} Free
response functions  for  ISGDR calculated using radial mesh size  $dr =
(0.04, 0.04)$  with
$\Gamma/2 = 0.025$ MeV and $\eta=24.51$ fm$^2$. The long dashed curve
clearly manifests the existence of the spuriocity over the entire range
of excitations but  predominant for the $1\hbar\omega$ region ($E <
20$ MeV).}

\vspace*{0.25 true in}
\caption{\label{isgdr-rpa} Same  as Fig. 2 but for fully self-consistent HF$-$CRPA results. 
The response for the operator $f_3$ and
$f_\eta$ is almost the same due to no spurious state mixing.}

\vspace*{0.25 true in}
\caption{\label{isgmr}  Free and self-consistent  HF$-$CRPA  results for the ISGMR strength function
calculated using radial mesh  $dr = (0.04, 0.04)$, $\Gamma/2 = 0.025$ MeV.}

\vspace*{0.25 true in}
\caption{\label{isgqr}  Same as Fig. 4 but for ISGQR.}

\vspace*{0.25 true in}
\caption{\label{isgdr-gam2}  Strength functions for the spurious state
and ISGDR   calculated using radial mesh size of 0.04 fm and smearing
parameter $\Gamma/2 = 1$ MeV in CRPA. The  SSM caused  due to long tail of spurisous state is
projected out using the operator $f_\eta$.}

\vspace*{0.25 true in}
\caption{\label{isgdr-drpa} Discretised RPA results for ISGDR response
obtained using different  values of the smearing parameter (a) $\Gamma/2
= 0.25$ MeV and  (b) $\Gamma/2 = 1.0$ MeV. The
discretization is performed using $N_{\scriptstyle{HF}} = 150$ (dotted line)
and $N_{\scriptstyle{HF}} = 900$ (solid line) with $dr = (0.08, 0.24)$. 
 We use particle-hole  cut-off energy $E_{ph}^{max} = 200$ MeV.} 

\vspace*{0.25 true in}
\caption{\label{rhoss-drpa} Comparison of the spurious state transition density obtained using
discretized RPA and collective model (dotted line) . The dash-dot, long dash and solid curves represent the
DRPA results for  $N_{\scriptstyle{HF}}\> (E_{ph}^{max})$ = 150 (50 MeV), 900 (50 MeV), 900 (200 MeV),
respectively.  The values of transition density do not change significantly when
$N_{\scriptstyle{HF}}$ increased from 150 to 900, but, with increase in $E_{ph}^{max}$ the 
DRPA  results become closer  to the collective model results.}

\vspace*{0.25 true in}
\caption{\label{e0-e1-kph} The centroid energies $E_0$ and $E_1$ versus
$\sqrt{K_{nm}^\prime}$ for $^{80}Zr$. Here, $K_{nm}^\prime$ denotes
the nuclear matter incompressibility coefficient associated with the
parameters used in particle-hole interaction (see also the text).}

\vspace*{0.25 true in}
\caption{\label{isgdr-tsc} Non self-consistent CRPA results for the ISGDR strength functions for the operators $f_3$ and
$f_\eta$ calculated using $t_0 = -1620$ MeVfm$^3$, radial mesh  size   $dr = (0.04, 0.04)$ and $\Gamma/2 = 0.25$ MeV.  The strength
function for the operator $f_\eta$ is larger than that for the operator $f_3$ for a wide range of energy.}

\vspace*{0.25 true in}
\caption{\label{isgr-selfcv} Influence of violation of self-consistency due to  variation of $t_0$   by
$+5\%$ (dashed line ) and $-5\%$ (dotted line) on the strength function for (a)  ISGMR, (b) ISGDR and (c) ISGQR .
 Solid line represents the self-consistent result (i.e., $t_0$=-1800 MeVfm$^3$).}

\vspace*{0.25 true in}
\caption{\label{isgdr-selfcv-drpa1} Comparison of DRPA results for  ISGDR strength functions
obtained  for (a) $t_0 = -1800$, (b) $t_0=-1620$ and (c) $t_0 = -1980$ MeV fm$^3$.  
Numerical calculations  for all the cases  are performed using,
  $N_{\scriptstyle{HF}} = 150$,  $dr = (0.08, 0.24)$, $E_{ph}^{max} =
50$ and $\Gamma/2 = 0.25$
MeV.}

\vspace*{0.25 true in}
\caption {\label{isgdr-selfcv-drpa2} Same as Fig. \ref{isgdr-selfcv-drpa1}
but for $E_{ph}^{max} = 200$ MeV.}
\end{figure}

\newpage
\begin{table}
\caption{ \label{sp-energy} Hartree-Fock single particle energies (in MeV) for the bound states
in $^{80}Zr$ nucleus obtained with the interaction parameters $t_0=-1800$ MeVfm$^{3}$, $t_3=-12871$
MeVfm$^{4}$ and  $\alpha=1/3$ using the small  mesh size $dr = 0.04$ fm.}
\begin{tabular}{|c|c|c|c|c|c|c|c|c|c|}
Orbits & $0s$& $0p$& $0d$& $1s$& $0f$& $1p$& $0g$& $1d$& $2s$\\
\hline
Energy & -45.50& -39.14& -31.02& -26.74&  -21.42& -15.33& -10.59& -3.98& -2.62\\
\end{tabular}
\end{table}

\newpage
\begin{table}
\caption{\label{exact-ewsr}Values for density radial moments $\langle r^2\rangle$ and $\langle r^4\rangle$ in units of fm$^2$
and fm$^4$, respectively, together with the EWSR associated with the scattering operator
$r^nY_{\scriptstyle{LM}}$, in units of fm$^{(2n)}$MeV,
for different mesh size $dr$ (in fm) used in the  HF calculations.}
\begin{tabular}{|c|c|c|c|c|c|c|c|}
\multicolumn{1}{|c|}{}&
\multicolumn{1}{c|}{}&
\multicolumn{1}{c|}{}&
\multicolumn{5}{c|}{EWSR}\\
\cline{4-8}
\multicolumn{1}{|c|}{$\quad dr\quad $}&
\multicolumn{1}{c|}{$\langle r^2\rangle$}&
\multicolumn{1}{c|}{$\langle r^4\rangle$}&
\multicolumn{1}{c|}{$rY_{10}$}&
\multicolumn{1}{c|}{$r^3Y_{10}$}&
\multicolumn{1}{c|}{$(r^3-\eta r)Y_{10}$}&
\multicolumn{1}{c|}{$r^2Y_{00}$}&
\multicolumn{1}{c|}{$r^2Y_{20}$}\\
\hline
0.04 & 14.705 & 282.147 & 391.04&404545 & 169661&  7667& 19167\\
0.08 & 14.702& 282.008& 391.04&404346& 169553 & 7665 & 19163\\
0.24 & 14.676& 280.653& 391.04&402403 & 168441 & 7651 & 19129\\
\end{tabular}
\end{table}

\newpage
\begin{table}
\caption {\label{str-0404}
The energy weighted transition strengths ($S^{\scriptstyle{EW}}$)  of the free and fully
self-consistent HF$-$CRPA  for $^{80}Zr$ nucleus (in fm$^6$MeV) calculated using
$dr_{\scriptstyle{HF}}=dr_{\scriptstyle{RPA}}=0.04$ fm,
$N_{\scriptstyle{RPA}}=300$ with no smearing width
($\Gamma = 0$ MeV).}
\begin{tabular}{|r|r|r|r|r|}
\multicolumn{5}{|c|}{Free response}\\
\hline
\multicolumn{1}{|c|}{Energy}&
\multicolumn{1}{c|}{$S^{\scriptstyle{EW}}_{3}$}&
\multicolumn{1}{c|}{$-2\eta S^{\scriptstyle{EW}}_{13}$}&
\multicolumn{1}{c|}{$\eta^2S^{\scriptstyle{EW}}_{1}$}&
\multicolumn{1}{c|}{$S^{\scriptstyle{EW}}_{\eta}$}\\
\hline
10.832306 & $\>$ 87689&-221289& 139609&$\>\>$6009\\
11.352610 & $\>$ 47160& $\>$ -99851& $\>$ 52854&$\>\>\>$163\\
12.709777 & $\>$ 24341& $\>$ -37010&$\>$  14068&$\>\>$1399\\
17.437181 & $\>$ 48562&$\>$-64831& $\>$ 21638& $\>\>$5369\\
35.163326 & $\>$ 17114& -7514&$\>\>\>$ 825&$\>$  10425\\
36.520494 &  $\>$ 5034& -2123&$\>\>\>$ 224&$\>\>$   3135\\
15.0-18.0 &  $\>\>\>$ 465&$\>\>$  393& $\>\>$ 528& $\>\>$ 1386\\
18.0-100.0& 172707&$\>$-36767&   5009& 140949\\
100.0-150.0&  1256& $\>\>\>$-609& $\>\>\>\>$89&$\>\>\>$ 736\\
\hline
Total      &404328&-469601&  234844& 169571\\
\hline
\multicolumn{5}{|c|}{CRPA response}\\
\hline  
0.078606  &234852&-469709& 234857& $\>\>$0  \\
11.434169  &$\>\>$  4480&$\>\>\>\>\>\>\>\>$5 & $\>\>\>\>$-1 &  $\>\>$ 4484\\
12.965783  &$\>\>$  1984&$\>\>\>\>\>\>\>\>$7 & $\>\>\>\>$ 0 & $\>\>$  1991\\
15.0-18.0  & $\>\>$ 6087&     $\>\>\>\>\>\>$45 & $\>\>\>\>\>\>$0 &  $\>\>$ 6132\\
18.0-100.0 &156848&    $\>\>\>\>\>$-42 & $\>\>\>\>$ 2 & 156808\\
100.0-150.0&$\>\>\>$ 258&    $\>\>\>\>\>$-13 & $\>\>\>\>\>$1 &$\>\>\>$ 246\\
\hline
Total      &404509& -469707& 234859& 169661\\
\end{tabular}
\end{table}

\newpage
\begin{table}
\caption {\label{str-2424-0424}Fully self-consistent  HF$-$CRPA results for
the energy weighted transition strengths  (in fm$^6$MeV) for $\Gamma = 0$
MeV using different
mesh sizes (in fm) and $N_{\scriptstyle{RPA}} = 50$ .}
\begin{tabular}{|r|r|r|r|r|}
\multicolumn{5}{|c|} {$dr_{\scriptstyle{RPA}}=dr_{\scriptstyle{HF}}=0.24$}\\ 
\hline
\multicolumn{1}{|c|}{Energy}&
\multicolumn{1}{r|}{$S^{\scriptstyle{EW}}_{3}$}&
\multicolumn{1}{r|}{$-2\eta S^{\scriptstyle{EW}}_{13}$}&
\multicolumn{1}{r|}{$\eta^2S^{\scriptstyle{EW}}_{1}$}&
\multicolumn{1}{r|}{$S^{\scriptstyle{EW}}_{\eta}$}\\
\hline
$\>$0.714539&  232751& -465617& 232866&   0\\
11.483532&$\>\>$    4214&     -18 &      0&$\>\>$    4196\\
%12.449512&       7.7&      -0.2 &     -1.5&       5.9\\
13.138693&$\>\>$    2306&    -124&      2&$\>\>$    2184\\
15.0-18.0$\>$ &$\>\>$  5693&     263&      3&$\>\>$    5959\\
18.0-100 & 154096&    792&    11 & 154899\\
100-150  &$\>\>\>$   184&    -7   &    1  &$\>\>\>$   178\\
\hline
Total&     399244&-464711&  232883&167416\\
\hline
\multicolumn{5}{|c|} {$dr_{\scriptstyle{RPA}}=6dr_{\scriptstyle{HF}}=0.24$}\\ 
\hline
11.429694 &$\>\>$   4470&  43    &   0 &$\>\>$   4513\\
12.962171 &$\>\>$   1998&  -2     & -4 &$\>\>$   1992\\
15.0-18.0 &$\>\>$   6158& -43       &1 &$\>\>$   6116\\
18.0-100.0&  159022&-2693&45&  156374\\
100.0-150.0&$\>\>\>$     363& -126& 19 &$\>\>\>$    256\\
\hline
  Total    & 172011&-2821& 61 &169251\\
\hline
\multicolumn{5}{|c|} {$dr_{\scriptstyle{RPA}}=6dr_{\scriptstyle{HF}}=0.24$, $V_{sc}=
0.9916^a$}\\ 
\hline
 0.099965  &237622&-474392& 236771&      1\\
11.430431  &$\>\>$  4505&  -27 &  0 & 4478\\
12.959961  &$\>\>$  2025&    -20 &      0 &   2005\\
15.0-18.0  &$\>\>$  6288&   -157&      3&   6134\\
18.0-100.0 &159324&  -2992&     52 & 156384\\
100.0-150.0&$\>\>\>$   368&   -128&     19 &$\>\>\>$    259\\
\hline
Total     &410132& -477716& 236845&  169260\\
\end{tabular}
$^{a)}$ Normalization of the particle-hole interaction to  put the spurious state at 0.1 MeV.

\end{table}

\newpage
\begin{table}
\caption{\label{centroids}HF based  CRPA results for the spurious state energy $E_{ss}$
and  centroid energy $E_{\scriptstyle{L}}$ for the ISGMR ($L = 0$), ISGDR ($L = 1$) and
ISGQR ($L = 2$)  (in MeV) 
obtained using $\Gamma/2 = 0.025$ MeV. For $L = 0$ and 2 resonances we use the energy range  $0 - 80$ MeV and for $L = 1$
we use $28 - 80$ MeV.}
\begin{tabular}{|c|c|l|c|c|c|c|}
\multicolumn{1}{|c|}{$\quad dr_{hf}\quad $}&
\multicolumn{1}{c|}{$dr_{rpa}$}&
\multicolumn{1}{l|}{$V_{sc}$}&
\multicolumn{1}{c|}{$E_{ss}$}&
\multicolumn{1}{c|}{$E_{0}$}&
\multicolumn{1}{c|}{$E_{1}$}&
\multicolumn{1}{c|}{$E_{2}$}\\
\hline
0.04 & 0.04& 1.0& 0.08 & 22.98& 35.88& 14.67 \\
0.08 & 0.08& 1.0& 0.18 & 22.97 & 35.86& 14.70\\
0.24& 0.24& 1.0&  0.71&  22.92& 35.80& 14.69 \\
0.04& 0.24& 1.0& $--^{*)}$ & 22.94& 35.83& 14.60 \\
0.04& 0.24& 0.9916& 0.09 & 22.98  & 35.85& 14.70 \\
0.04& 0.24&  0.9707& 2.00& 23.08 & 35.88& 14.96 \\
\end{tabular}
$^{*)}$$E_{ss}$ is imaginary.
\end{table}

\newpage
\begin{table}
\caption{\label{ewsr} HF$-$CRPA results for fraction  energy weighted
sum rule (in percent)  for the spurious state (SS) and
for $L = 0 - 2$ resonances calculated using  various radial mesh sizes
$dr_{\scriptstyle{HF}}$ and $dr_{\scriptstyle{RPA}}$ (in fm) and the
energy region 0 - 80 MeV for $\Gamma/2=0.025$ MeV$^*$. See Table V for the
corresponding values of $E_{ss}$.}
\begin{tabular}{|c|c|c|c|c|c|c|}
%\multicolumn{1}{|c|}{}&
%\multicolumn{1}{c|}{}&
%\multicolumn{1}{c|}{}&
%\multicolumn{4}{c|}{$\%$EWSR }\\
%\cline{4-7}
\multicolumn{1}{|c|}{$\quad dr_{HF}\quad $}&
\multicolumn{1}{c|}{$dr_{RPA}$}&
\multicolumn{1}{c|}{$V_{scale}$}&
\multicolumn{1}{c|}{SS}&
\multicolumn{1}{c|}{$L = 0$}&
\multicolumn{1}{c|}{$L = 1$}&
\multicolumn{1}{c|}{$L = 2$}\\
\hline
0.04 & 0.04& 1.0& 99.99& 99.84& 99.61& 99.91\\
0.08 & 0.08& 1.0& 99.95 & 99.76 & 99.76& 99.91 \\
0.24& 0.24& 1.0&  99.55&  99.74&99.25& 99.49 \\
0.04& 0.24& 1.0& $--$ & 102.05 & 99.57& 101.18 \\
0.04& 0.24& 0.9916& 101.22 & 102.02  & 99.57& 101.17 \\
0.04& 0.24&  0.9707& 101.58& 102.96 & 99.34& 101.15 \\
\end{tabular}
$^*$For the spurious state we use $\Gamma  = 0$ and Eq.(8).
\end{table}

\newpage
\begin{table}
\caption{\label{ewsr-isgdr} CRPA results for the fraction  energy weighted sum
rule (in percent)   of the ISGDR  obtained using the operator $f_\eta$ for the energy
range $\omega_1-\omega_2$   (in MeV) for various combinations
 of the mesh size (in fm)  and smearing parameter $\Gamma/2$ (in MeV). } 
\begin{tabular}{|c|c|c|c|c|c|c|c|}
\multicolumn{1}{|c|}{}&
\multicolumn{1}{c|}{}&
\multicolumn{1}{c|}{}&
\multicolumn{4}{c|}{$\omega_1 - \omega_2$}&
\multicolumn{1}{c|}{}\\
\cline{4-7}
\multicolumn{1}{|c|}{$\quad dr_{HF}\quad $}&
\multicolumn{1}{c|}{$dr_{RPA}$}&
\multicolumn{1}{c|}{$\Gamma/2$}&
\multicolumn{1}{c|}{ 0 - 15}&
\multicolumn{1}{c|}{ 15 - 18}&
\multicolumn{1}{c|}{ 18 - 100}&
\multicolumn{1}{c|}{ 100 - 150}&
\multicolumn{1}{c|}{ Total}\\
\hline
0.04&  0.04&  0.0$\>\>\>\>$&  3.82&  3.61&  92.42&  0.15&  100.00$\>\>$\\
0.04&  0.04&  0.025&  3.81&  3.59&  92.40&  0.16&  99.96\\
0.04&  0.04&  0.25$\>\>$&      3.79  & 3.33 & 92.38  & 0.27 & 99.77\\
0.04&  0.04&  1.0$\>\>\>\>$&      3.69 &  2.89 & 91.87  & 0.65 & 99.10\\
0.24&  0.24&  0.0$\>\>\>\>$&    3.79&  3.54&  91.96&  0.11&  99.40\\
0.24&  0.24&  0.025&  3.78  & 3.51&  91.95  & 0.12 & 99.36\\
0.24&  0.24&  0.25$\>\>$&         3.75  & 3.29 & 91.91 &  0.23 & 99.18\\
0.24&  0.24&  1.0$\>\>\>\>$&     3.63 &  2.88 & 91.39 &  0.61 & 98.51\\
0.04&  0.24&  0.0$\>\>\>\>$&    3.83&  3.60&  92.17&  0.15&  99.75\\
0.04&  0.24&  0.025 &3.83  & 3.48&  92.16 &  0.16 & 99.63\\
0.04&  0.24&  0.25$\>\>$ &3.80  & 3.33 & 92.13  & 0.28 & 99.54\\
0.04 &  0.24&  1.0$\>\>\>\>$&       3.71 &  2.89 & 91.62  & 0.65 & 98.87\\
%\multicolumn{8}{|c|}{$V_{sc}=0.99165$, $E_{ss}=0.100$}\\
%\hline
0.04$^*$&  0.24&  0.0$\>\>\>\>$&    3.82& 3.61&  92.17&  0.15&  99.75\\
0.04$^*$&  0.24&  0.025&  3.82 & 3.49 & 92.17 &  0.16 & 99.64\\
0.04$^*$&  0.24&  0.25$\>\>$& 3.79 &  3.34 & 92.14  & 0.28 & 99.55\\
0.04$^*$&  0.24&  1.0$\>\>\>\>$&        3.69 &  2.90&  91.63&   0.66 & 98.88\\
\end{tabular}
\end{table}
$^* V_{sc} = 0.9916$ and $E_{ss} = 0.1$ MeV.

\newpage
\begin{table}
\caption{\label{ewsr-drpa} HF$-$DRPA results for $E_{ss}$ and   the fraction of
energy weighted sum rule  of the ISGDR obtained using $f_\eta$ (in percent)
in the energy range $\omega_1 - \omega_2$ for various combinations of
$N_{HF}$, $E_{ph}^{max}$ and $\Gamma/2$ with $V_{sc}=1.0$, $N_{RPA}=50$,
$dr_{HF}=0.08$ fm and $dr_{RPA}=0.24$ fm.  Values  of $\omega$,  $E_{ss}$,
$E_{ph}^{max}$ and $\Gamma/2$ are in MeV.}
\begin{tabular}{|c|c|c|c|c|c|c|c|c|}
\multicolumn{1}{|c|}{}&
\multicolumn{1}{c|}{}&
\multicolumn{1}{c|}{}&
\multicolumn{1}{c|}{}&
\multicolumn{4}{c|}{$\omega_1 - \omega_2$}&
\multicolumn{1}{c|}{}\\
\cline{5-8}
\multicolumn{1}{|c|}{$\quad N_{HF}\quad $}&
\multicolumn{1}{c|}{$E_{ph}^{max}$}&
\multicolumn{1}{c|}{$\Gamma/2$}&
\multicolumn{1}{c|}{$E_{ss}$}&
\multicolumn{1}{c|}{ 0 - 15}&
\multicolumn{1}{c|}{ 15 - 18}&
\multicolumn{1}{c|}{ 18 - 100}&
\multicolumn{1}{c|}{ 100 - 150}&
\multicolumn{1}{c|}{ Total}\\
\hline
150&  50 & 0.25& 4.4& 3.65&  2.74& 88.74&  0.00&   $\>\>$95.13\\
150&  200&  0.25&1.3&  3.81&  3.07&  93.22&  0.42&   100.52\\
150&  400&  0.25&$--$& 3.84&  3.07&  93.22&  0.41&  100.54\\
150&  50 & 1.0$\>\>$& 4.3&  3.71&  2.80&  85.71&  0.00& $\>\>$92.22\\
150&  200&  1.0$\>\>$&1.1&   3.89&  3.02&  93.29&  1.21& 101.41\\
150&  400&  1.0$\>\>$& $--$&  3.92&  3.03&  93.26&  1.21&  101.42\\
900&  50 & 0.25& 4.7& 3.64&  3.11&  85.01& 0.00&  $\>\>$91.77\\
900&  200&  0.25& 1.5&   3.79&  3.43&  90.68&  0.44& $\>\>$98.34\\
900&  400&  0.25& 1.0&   3.82&  3.43& 90.67& 0.44& $\>\>$98.36\\
900&  50 & 1.0$\>\>$& 4.6&   3.70&  2.80&  82.82&  0.00& $\>\>$89.32\\
900&  200&  1.0$\>\>$& 1.4&   3.88&  3.03&  91.16&  1.22&  $\>\>$99.29\\
900&  400&  1.0$\>\>$& 0.7  & 3.90&  3.04&  91.15& 1.21&  $\>\>$99.30\\
\end{tabular}
\end{table}

\newpage
\begin{table}
\caption{\label{ecen-eph} Dependence of $E_{ss}$ and  the centroid
energies $E_L$ ($L = 0$, 1 and 2), in MeV,  on the value of $E_{ph}^{max}$ (in MeV) used
in HF$-$DRPA calculations. We have used the values of  $N_{\scriptstyle{HF}} = 900$,
$N_{\scriptstyle{RPA}} = 50$,  $dr = (0.08, 0.24)$ and $\Gamma/2 = 0.25$ MeV. The
corresponding HF$-$CRPA results  are placed in the last row.}
\begin{tabular}{|c|c|c|c|c|}
\multicolumn{1}{|l|}{$\>\>\> E_{ph}^{max}\>\>\>$}&
\multicolumn{1}{c|}{$E_{ss}$}&
\multicolumn{1}{c|}{$E_{0}$}&
\multicolumn{1}{c|}{$E_{1}$}&
\multicolumn{1}{c|}{$E_{2}$}\\
\hline
50& 4.7&23.92&35.34&16.11\\
75&3.3&23.51&35.76&15.51\\
100& 2.9&23.25&35.66&15.14\\
200&1.5&23.09&35.55&14.82\\
400&1.0&23.02&35.51&14.73\\
600 & 0.9 & 23.02 & 35.51& 14.72\\
$\infty$ &  0.7 & 23.01&  35.46& 14.70\\
\end{tabular}
\end{table}

\newpage
\begin{table}
\caption{ \label{vzz-vs-ecen} HF$-$CRPA results for the spurious state energy  $E_{ss}$, incompressiblity
coefficient $K^\prime_{nm}$  and centroid energy $E_L$ (in MeV) for isoscalar 
giant resonances for $L = 0 - 2$  with different values of $t_0$, $t_3$ and $V_{sc}$ used 
 in the particle-hole interaction.
These calculations are performed using $\Gamma/2 = 0.25$ MeV
and $dr_{\scriptstyle{HF}}=dr_{\scriptstyle{RPA}}=0.04$fm.}
\begin{tabular}{|c|c|l|c|c|c|c|c|}
\multicolumn{1}{|c|}{$\qquad t_0\qquad$}&
\multicolumn{1}{c|}{$\qquad t_3\qquad$}&
\multicolumn{1}{l|}{$V_{sc}$}&
\multicolumn{1}{c|}{$K_{nm}^\prime$}&
\multicolumn{1}{c|}{$E_{ss}$}&
\multicolumn{1}{c|}{$E_{0}$}&
\multicolumn{1}{c|}{$E_{1}$}&
\multicolumn{1}{c|}{$E_{2}$}\\
\hline
-1800 & 12871&  1.0& 226 & 0.1 & 23.1& 35.5& 14.8 \\
-1710 & 12871 & 1.0& 258 & 6.7 & 26.3 & 37.9 & 17.4\\
-1710 & 12871 & 1.2938& 321 & 0.1  & 26.0 & 38.2& 14.7\\
-1620 & 12871 & 1.0& 289 & 9.2 & 29.0& 40.0& 19.5\\
-1620 & 12871 & 1.7118& 464 & 0.1 & 29.8& 41.8 & 14.7\\
-1620 & 11875& 1.0 & 226& 5.9& 24.9& 36.7& 16.8 \\
-1620 & 11270& 1.0& 188& 0.1& 21.6& 34.4& 14.8\\
-1890 & 12871& 1.0& 194 & $--$ & 18.7& 32.8& 11.1\\
-1890 & 12871 & 0.7910& 163 & 0.1 & 20.8& 33.7&  14.8\\
-1980 & 12871 & 1.0 & 162 & $--$ & 11.4& 29.9& 2.1\\
-1980 & 12871 & 0.6398& 120 & 0.1 & 19.2& 32.6& 14.9 \\
-1980& 13875& 1.0 & 226& $--$ & 20.8& 34.2& 12.1\\
-1980& 14500& 1.0& 266& 0.1& 24.3& 36.6& 14.7\\
\hline
\end{tabular}
\end{table}
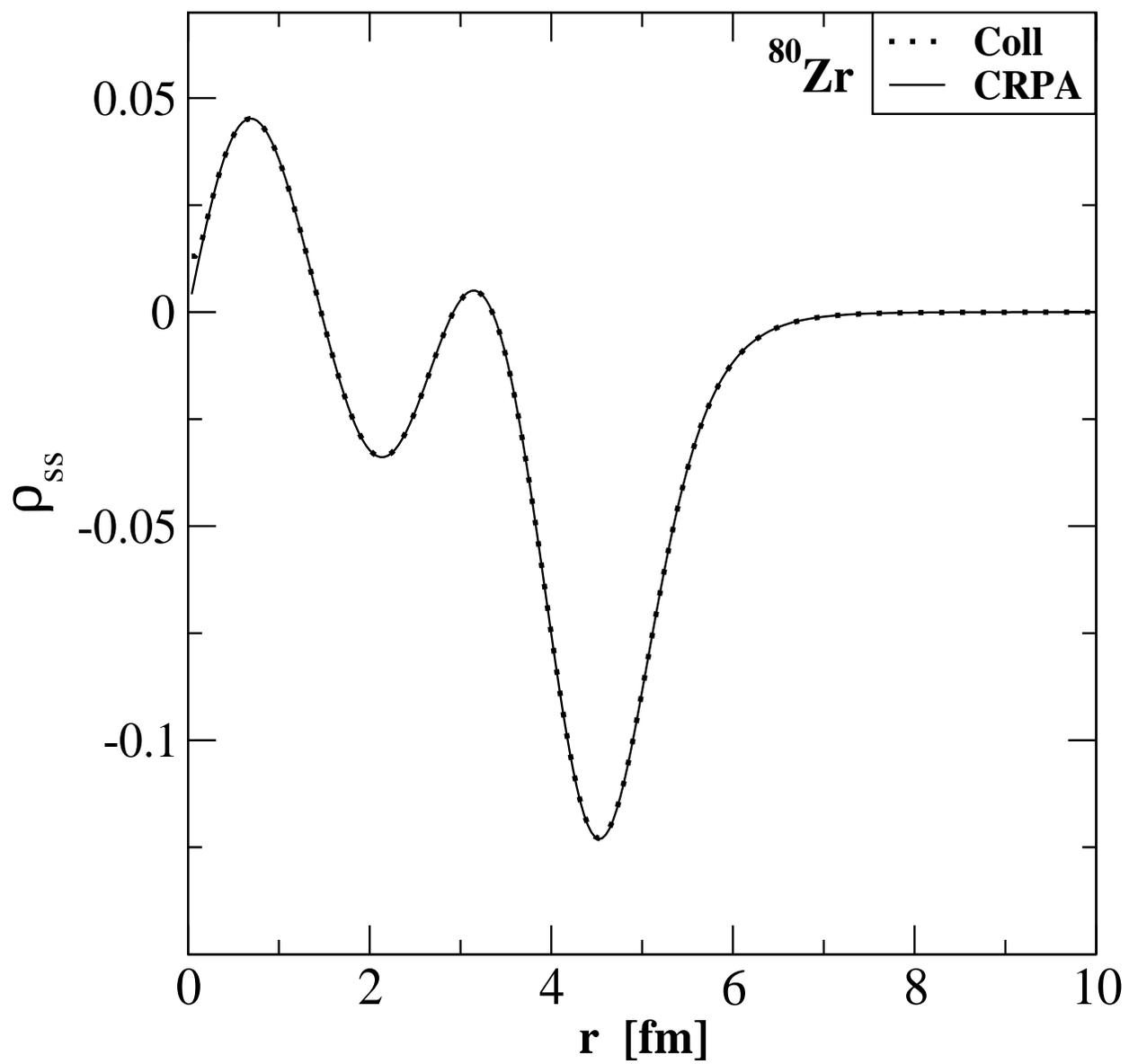
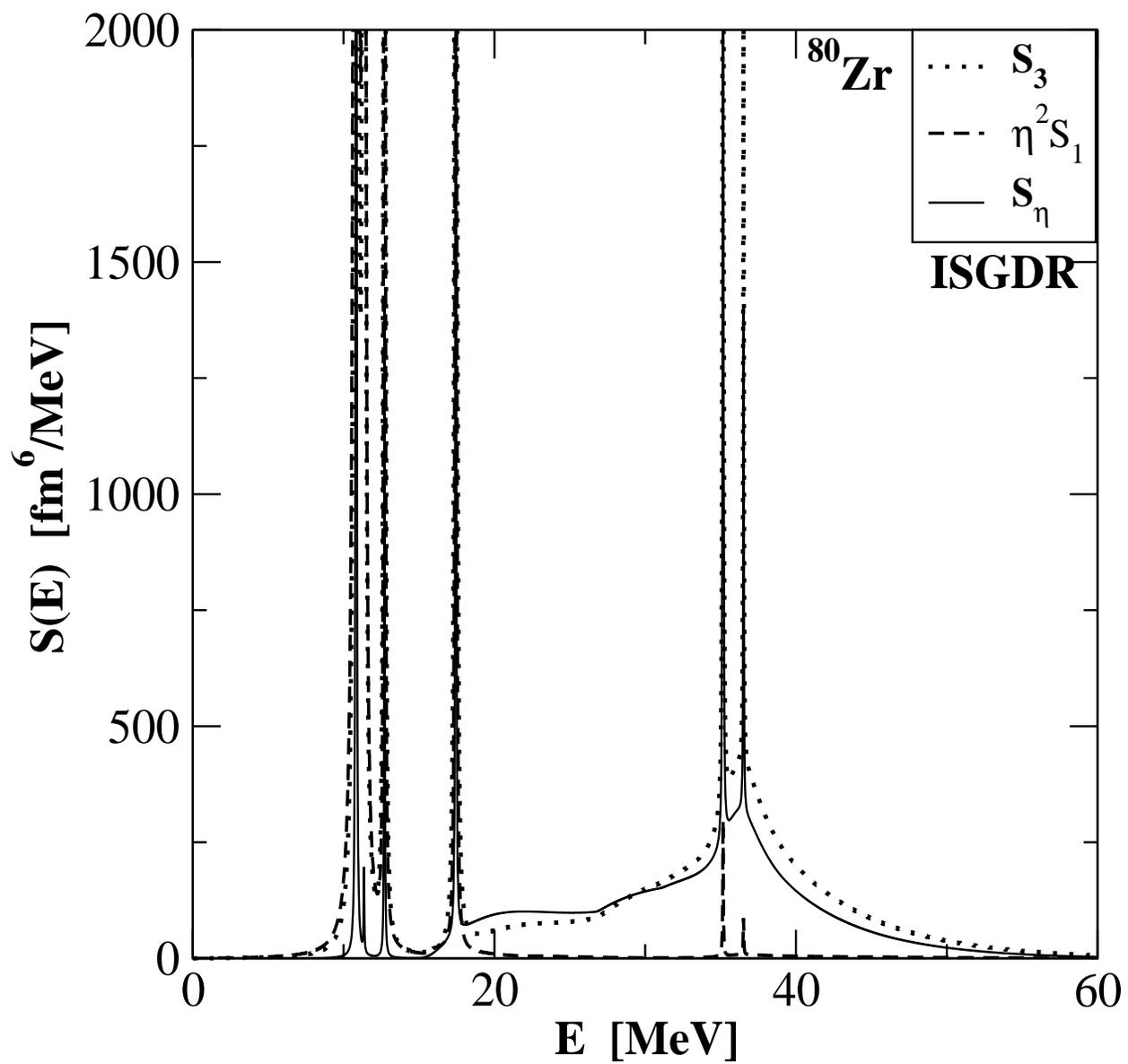
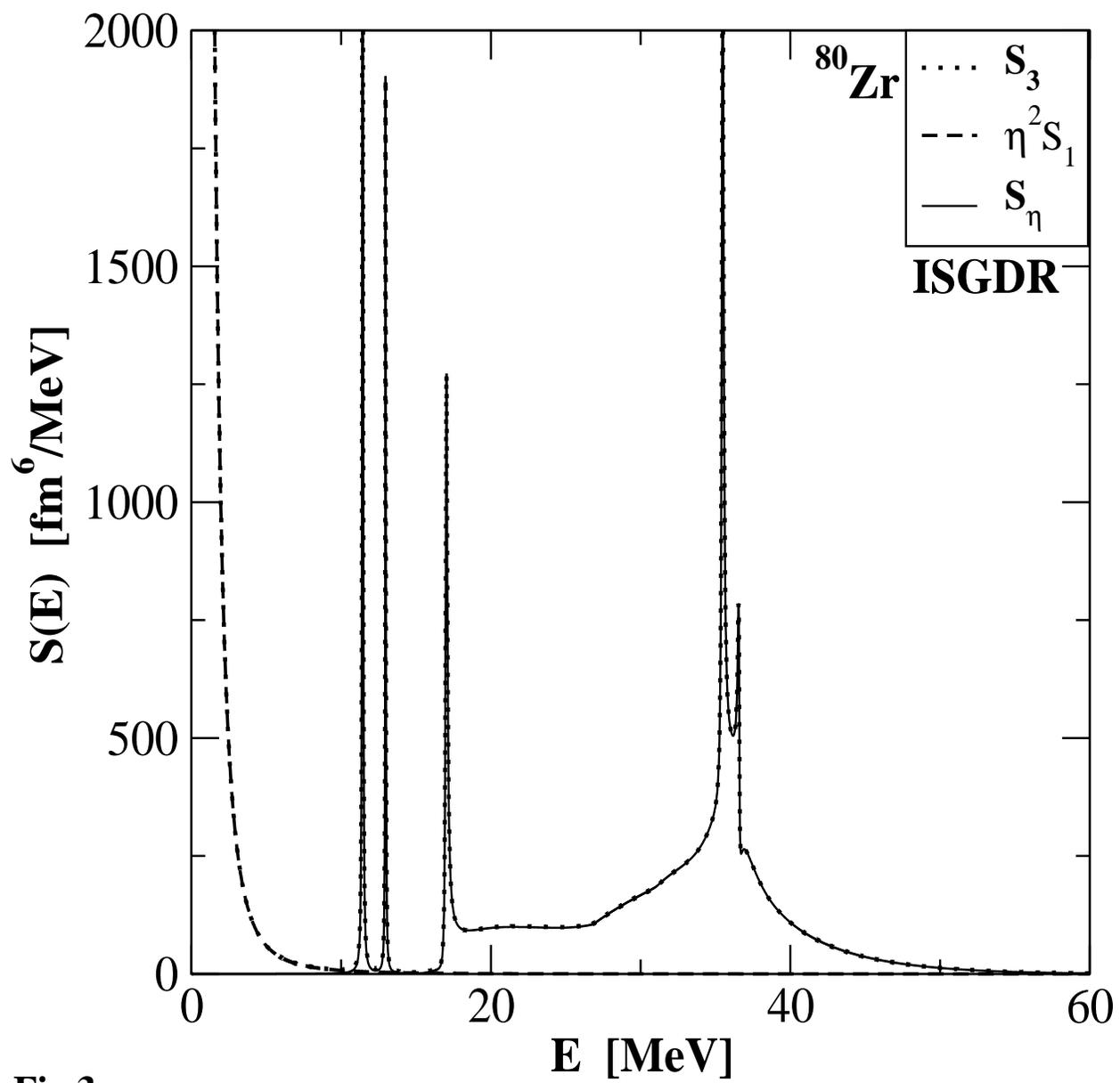
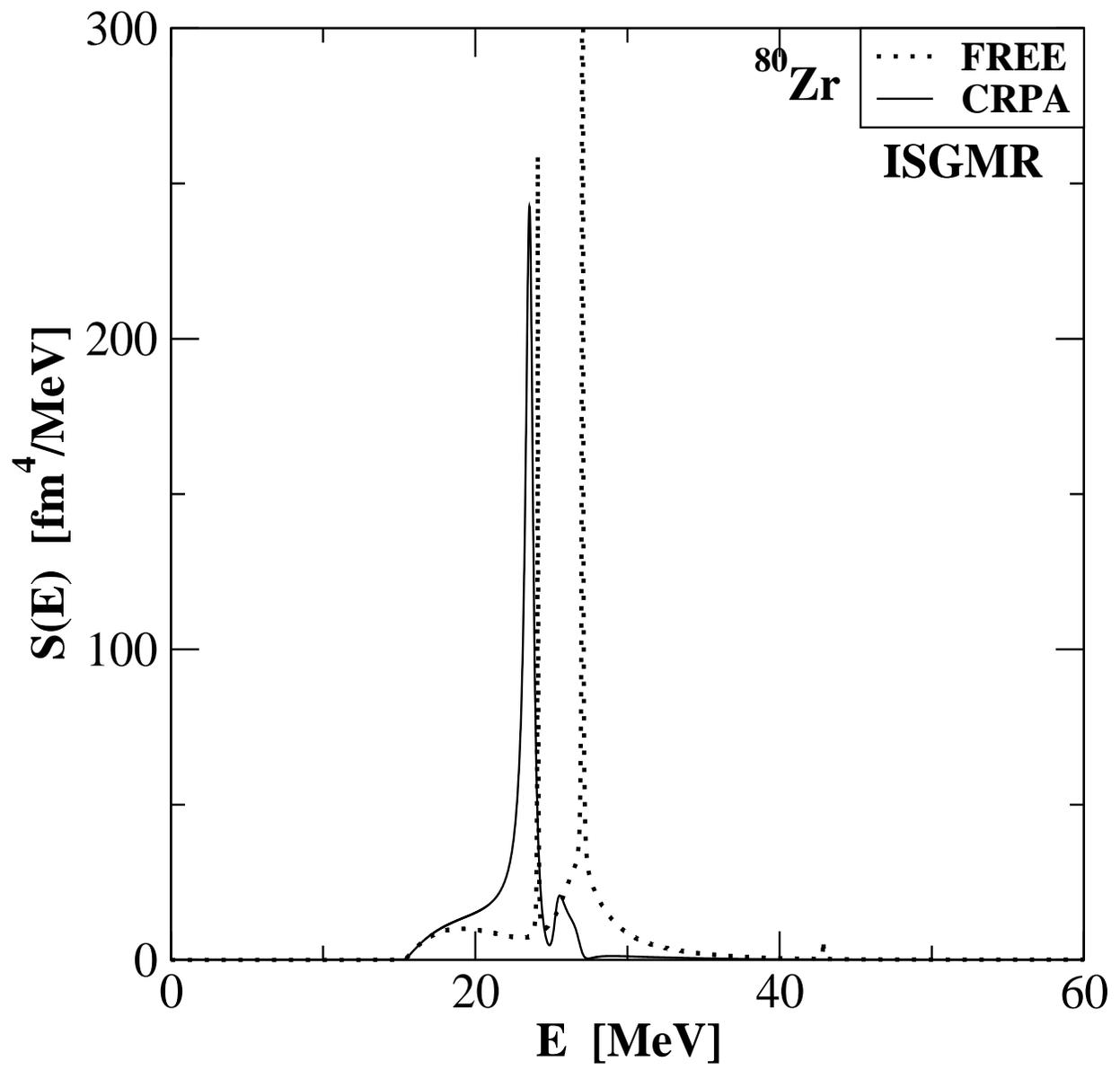
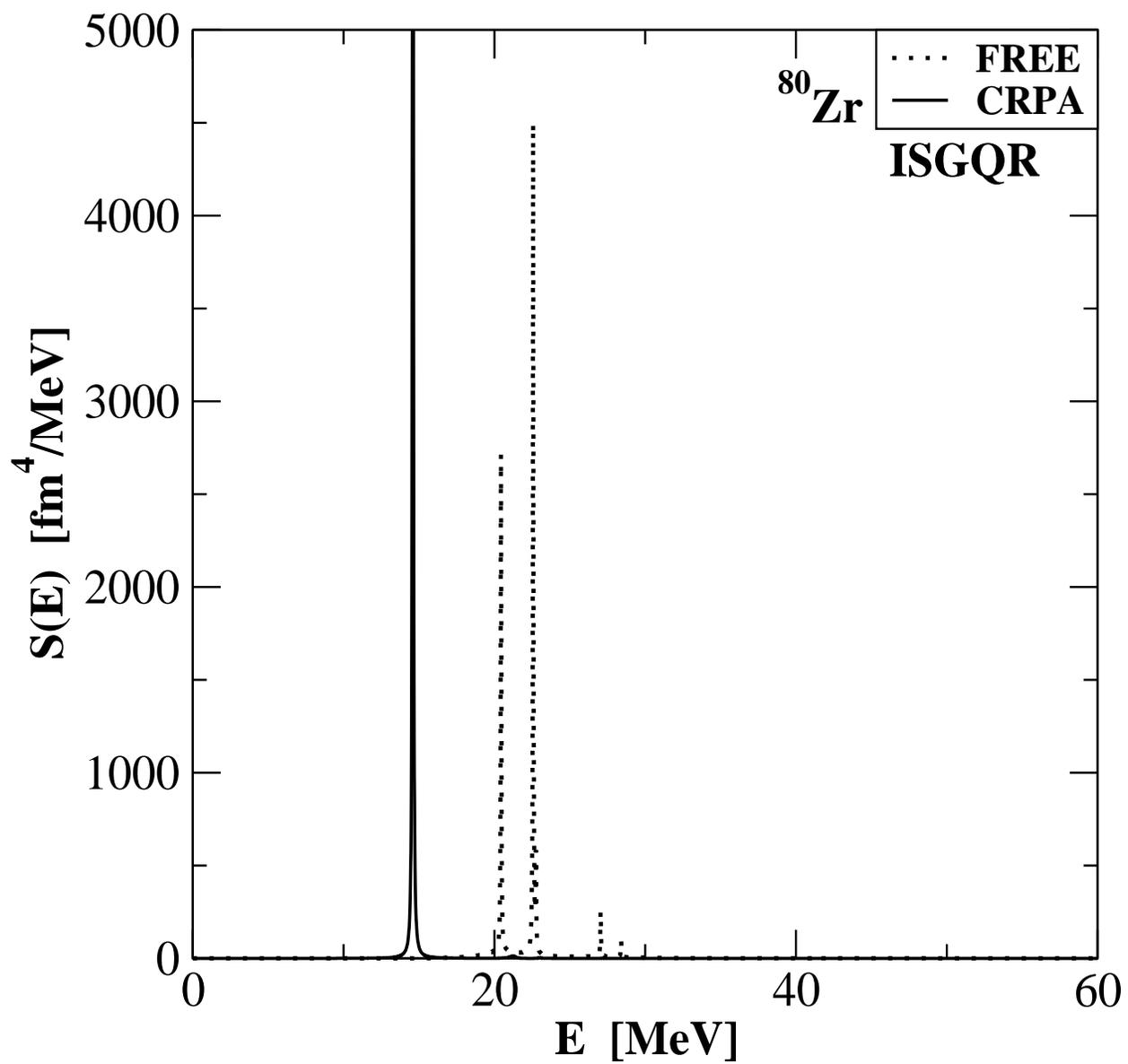
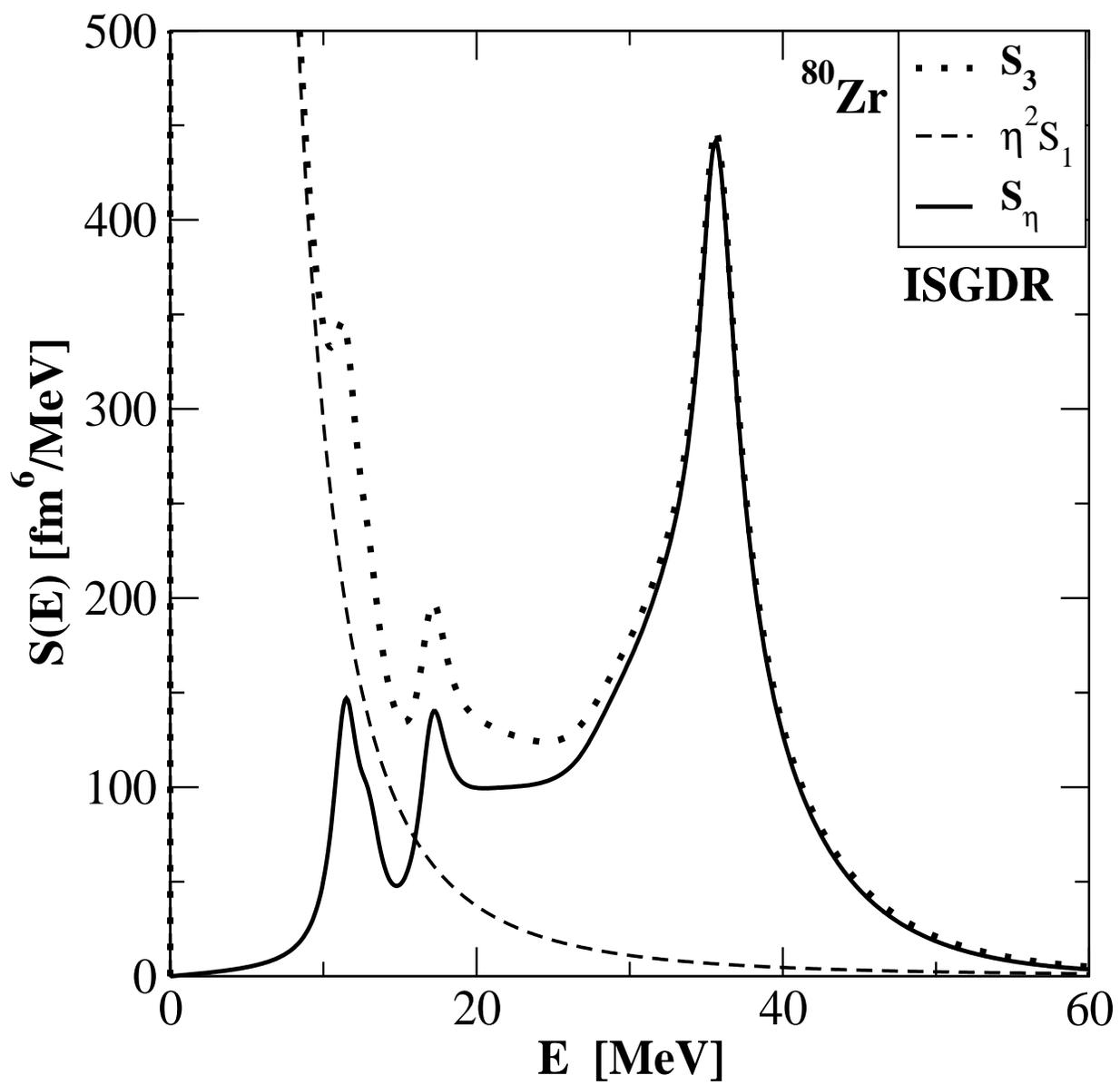
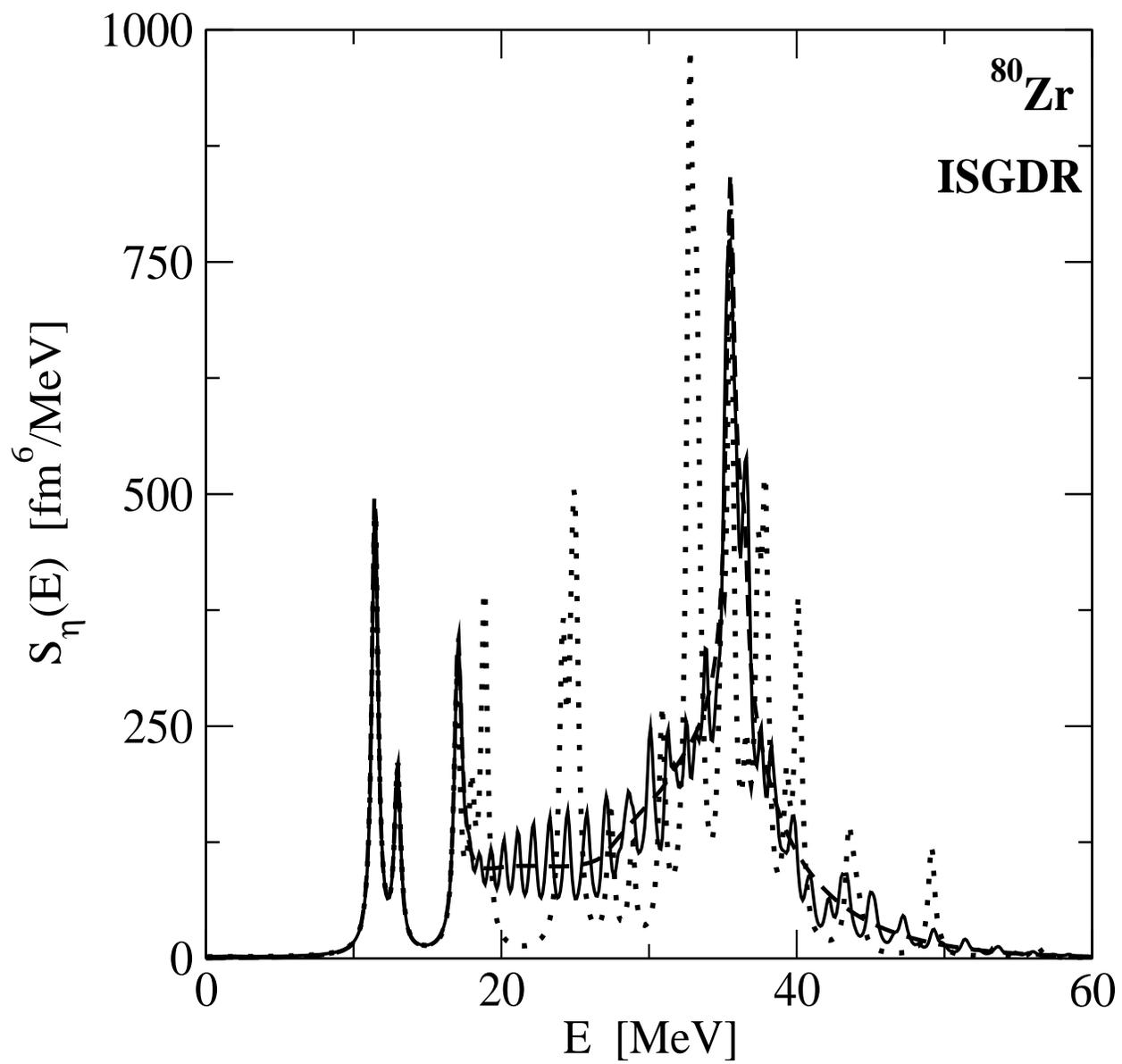
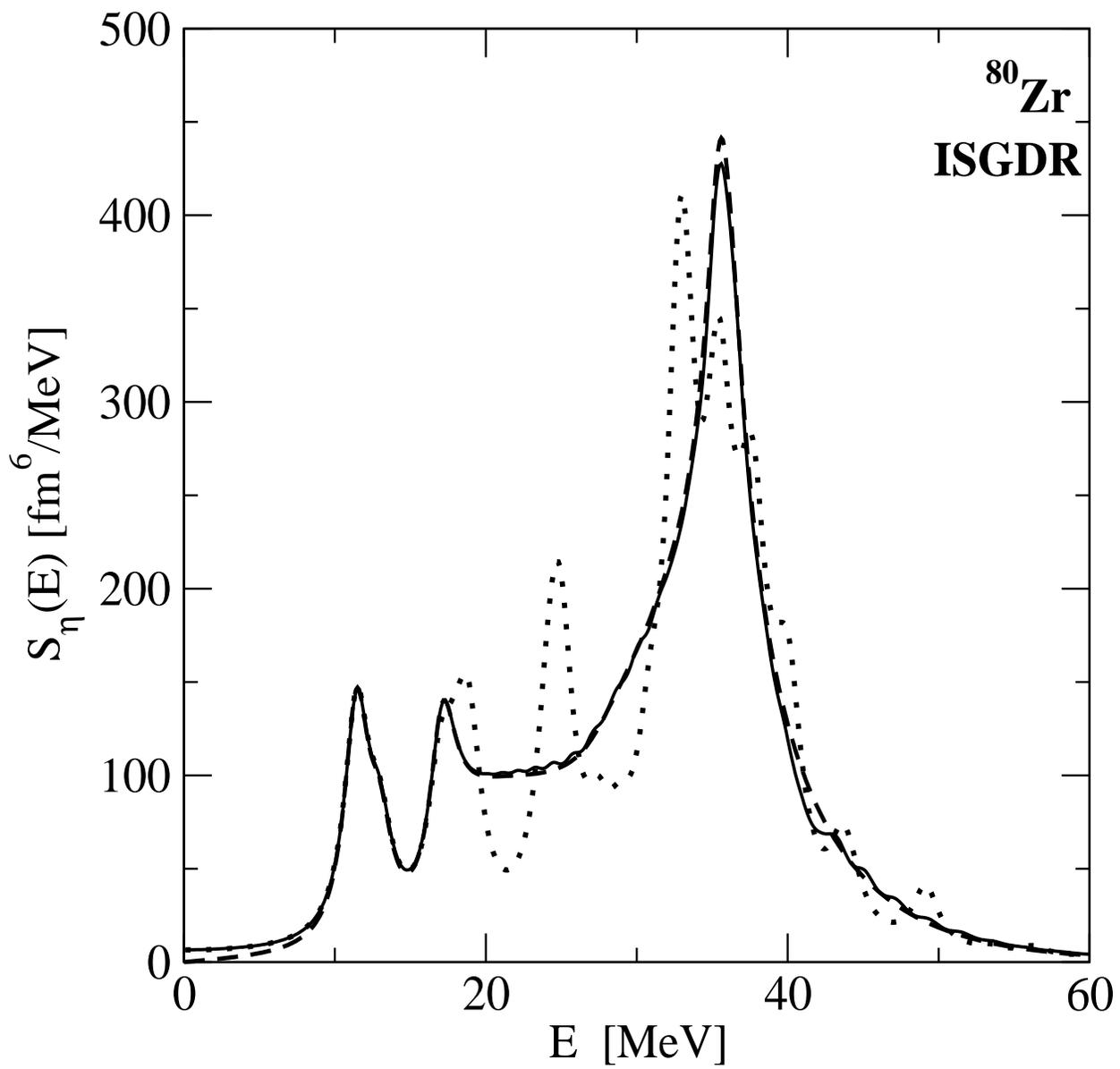
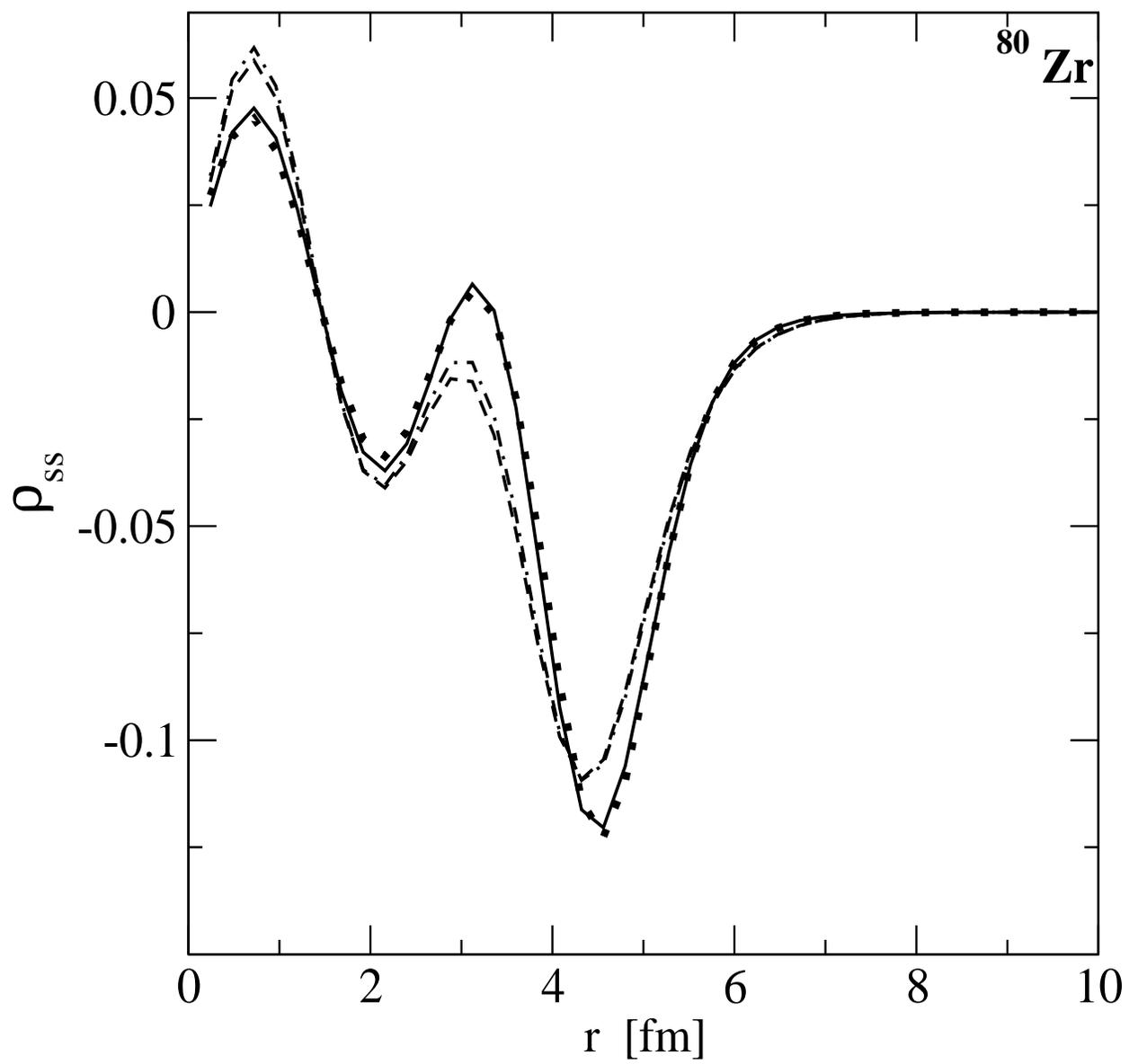
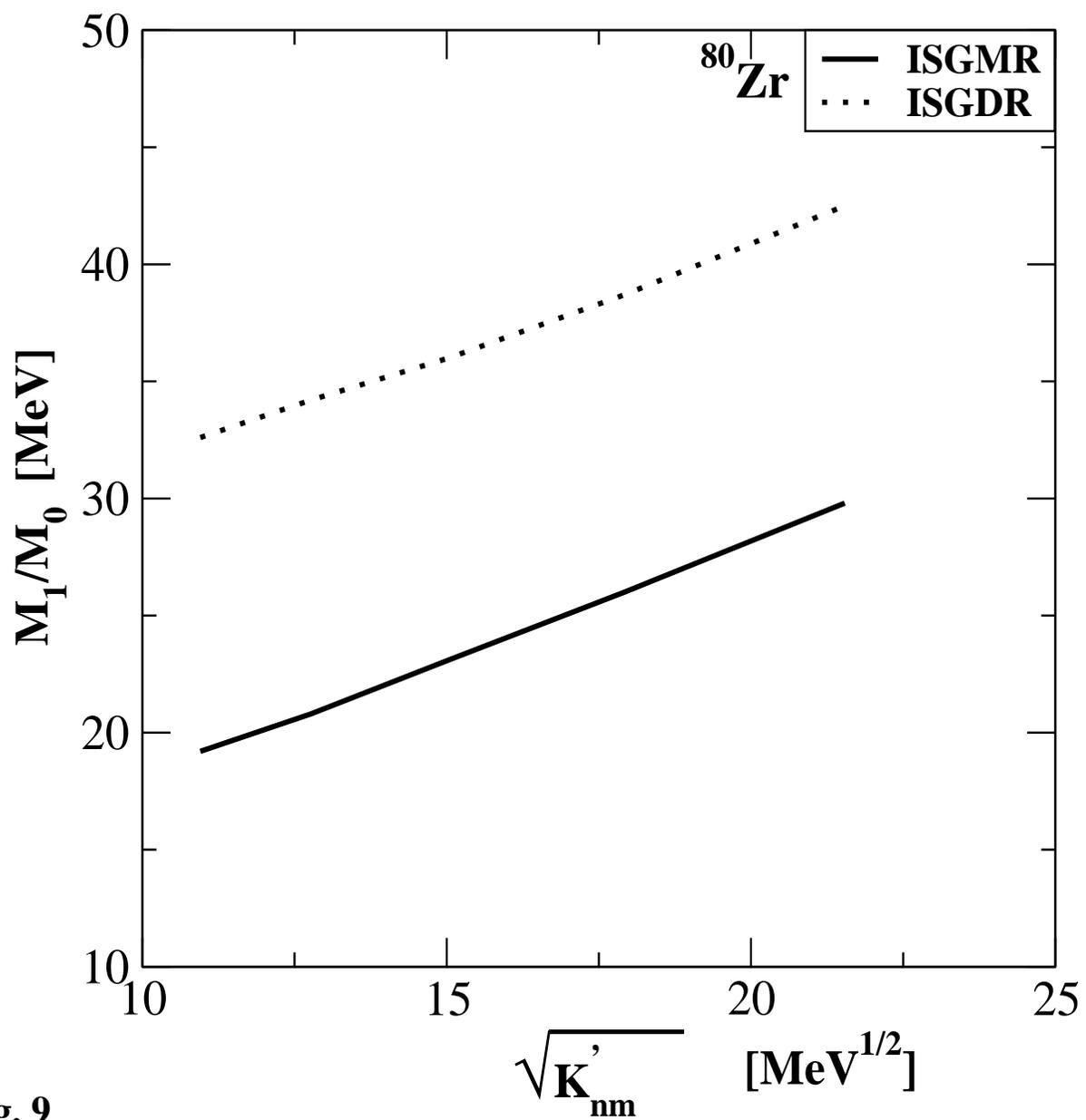
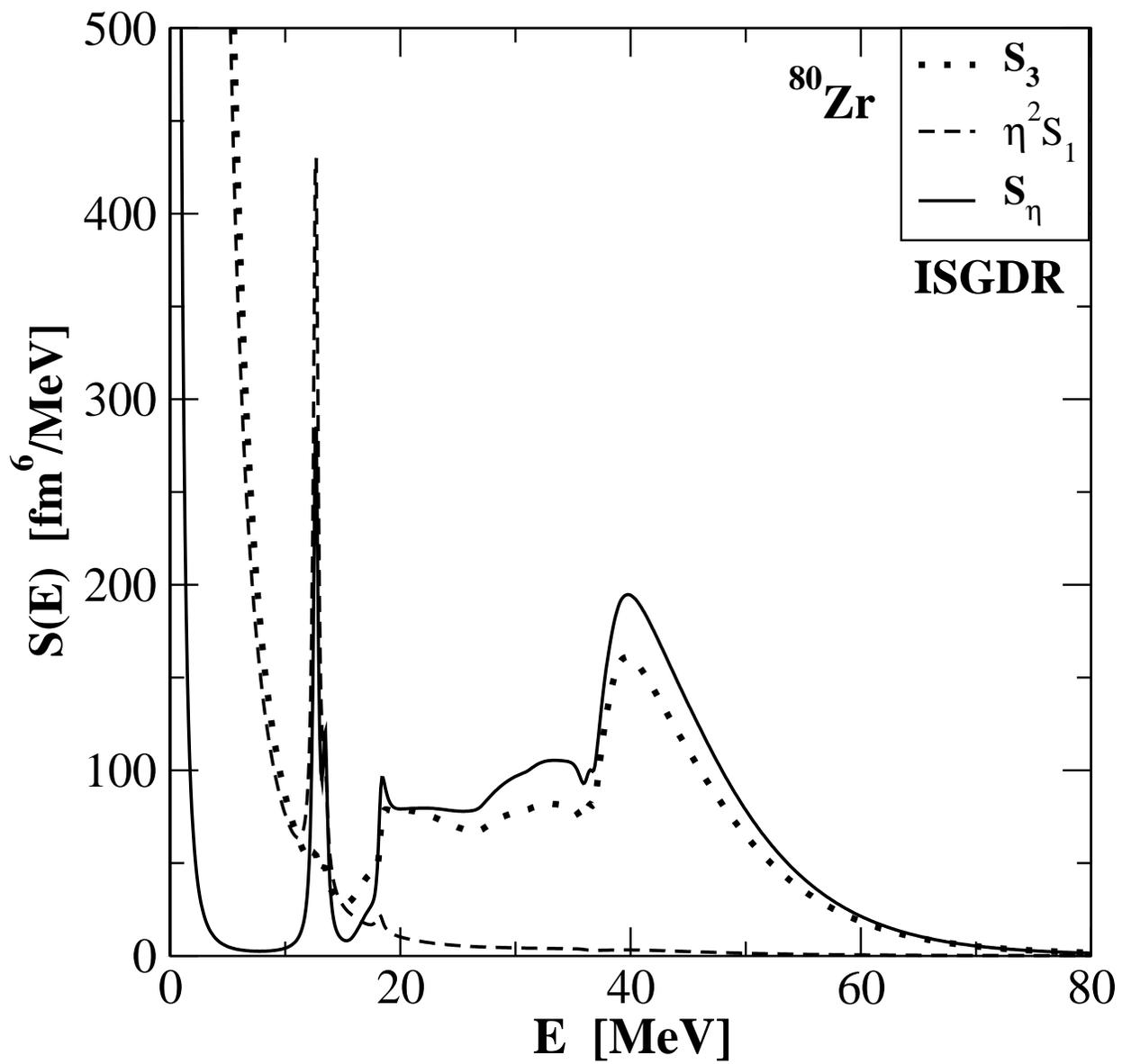
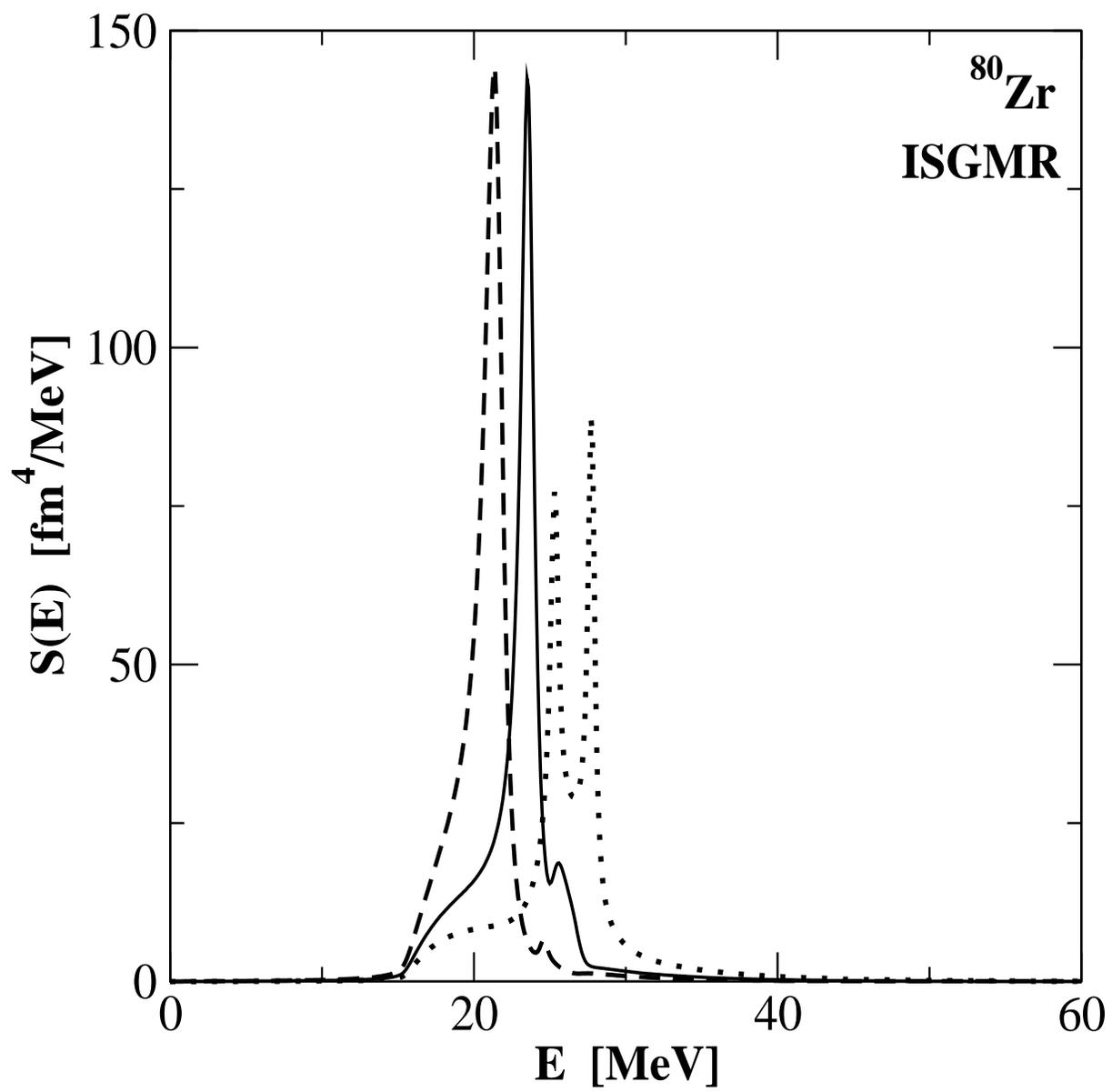
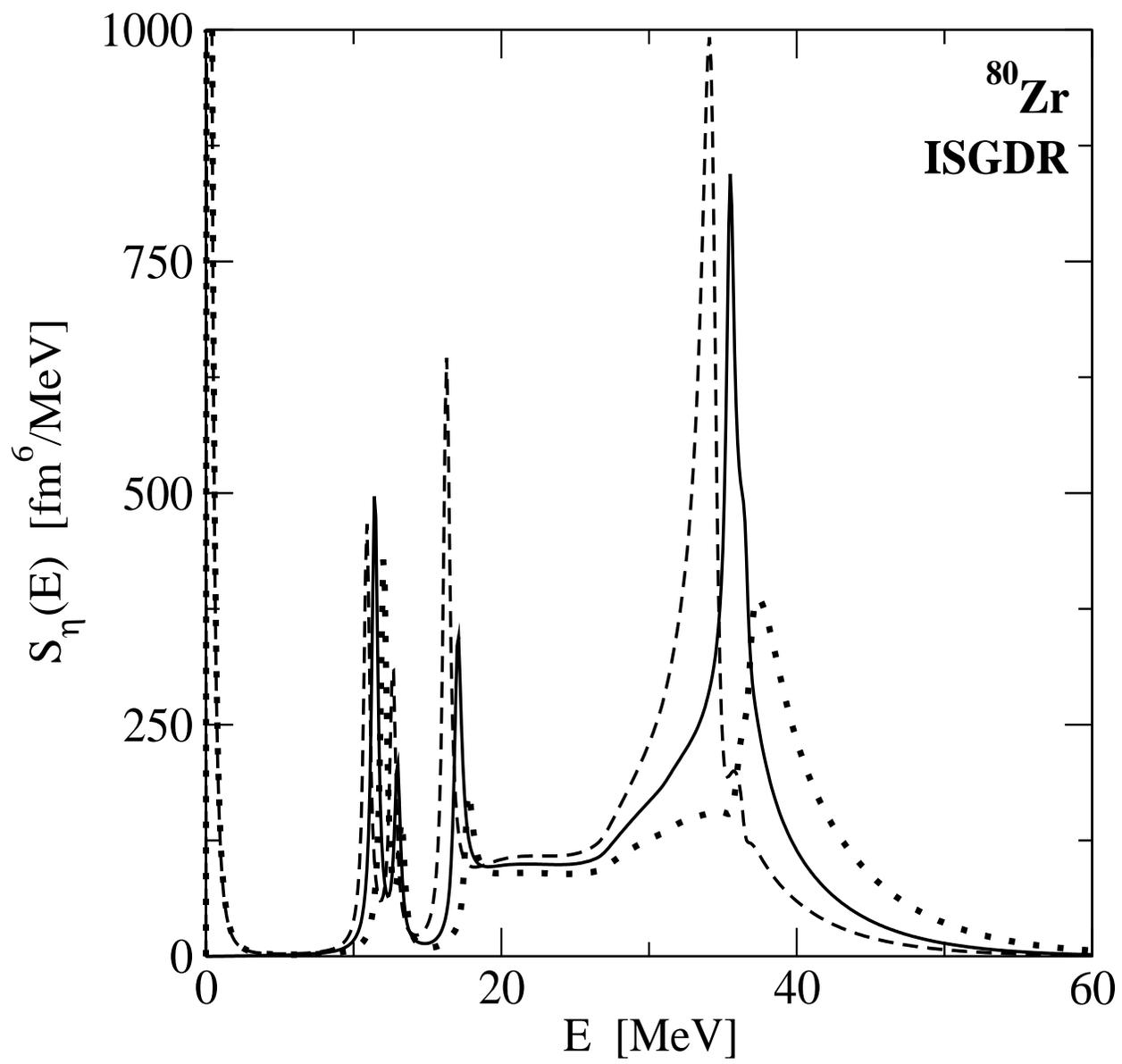
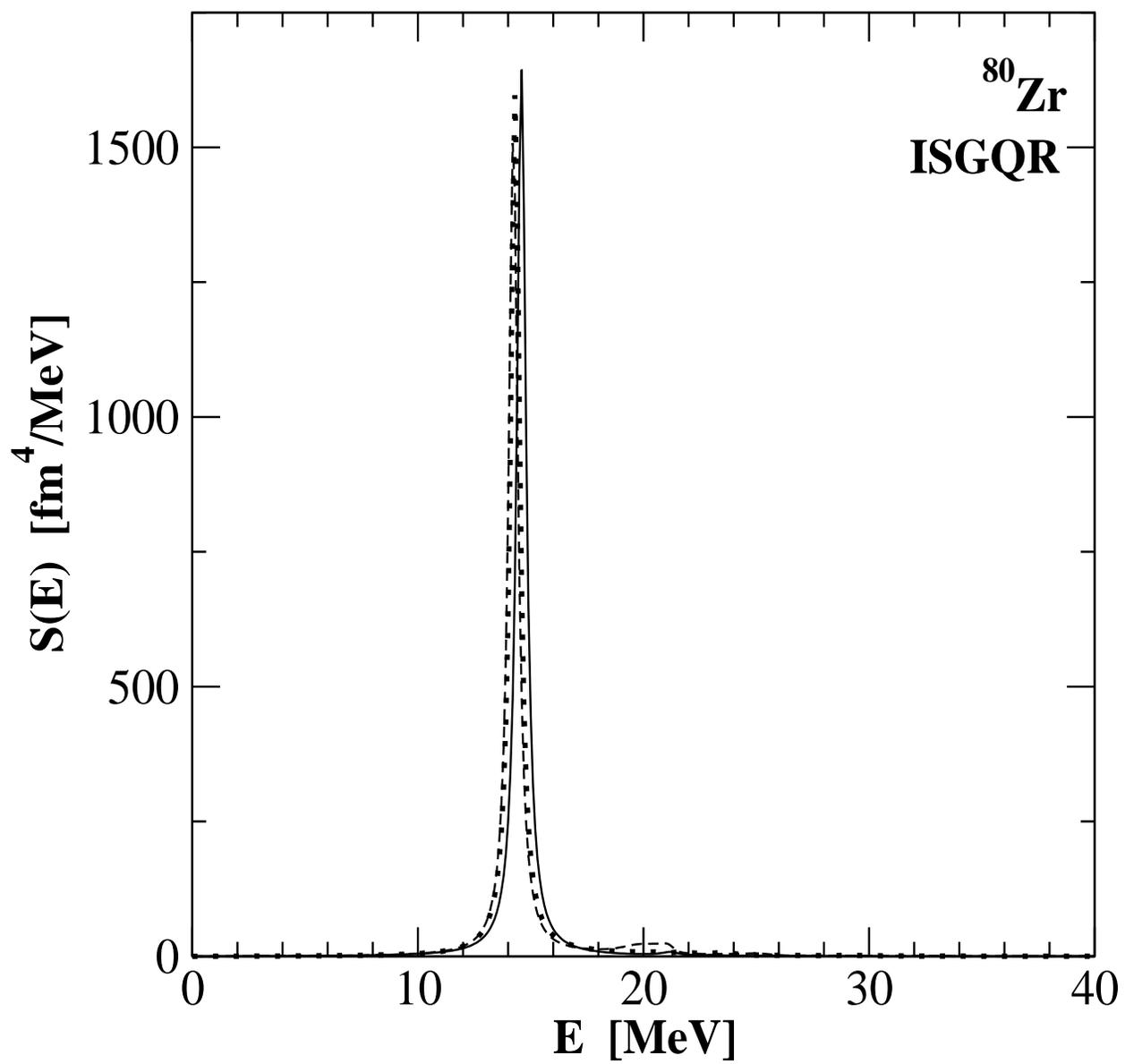
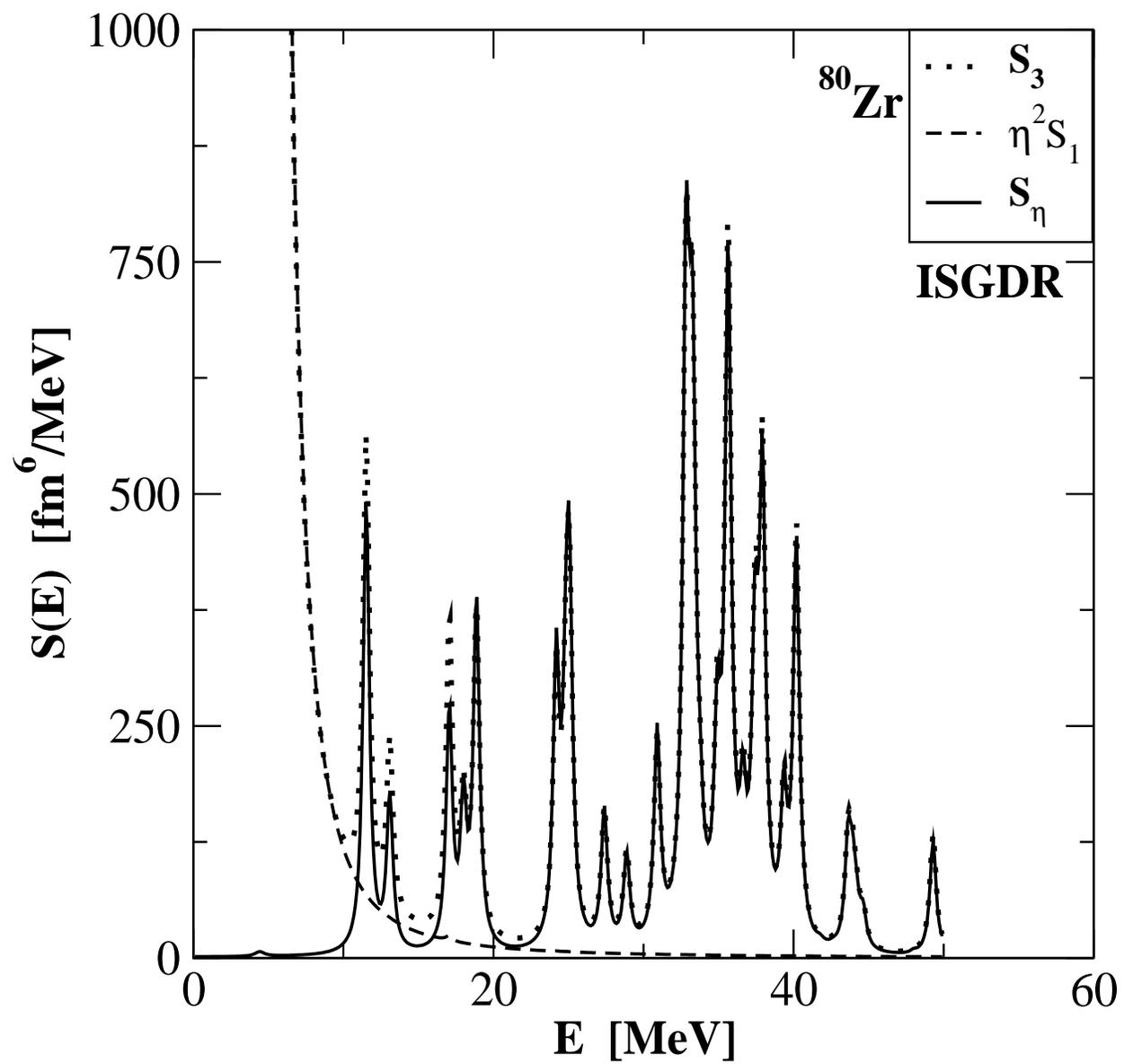
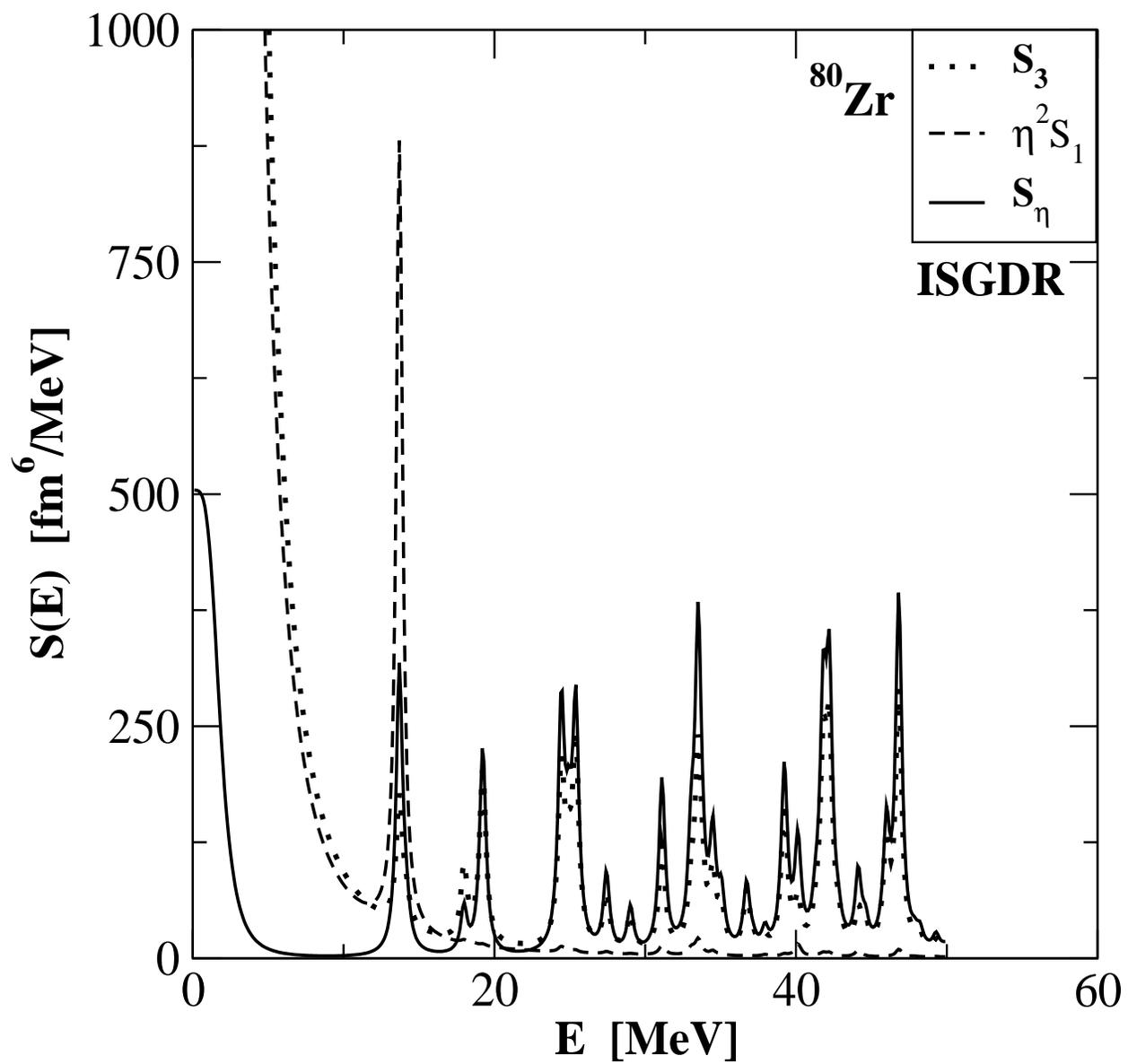
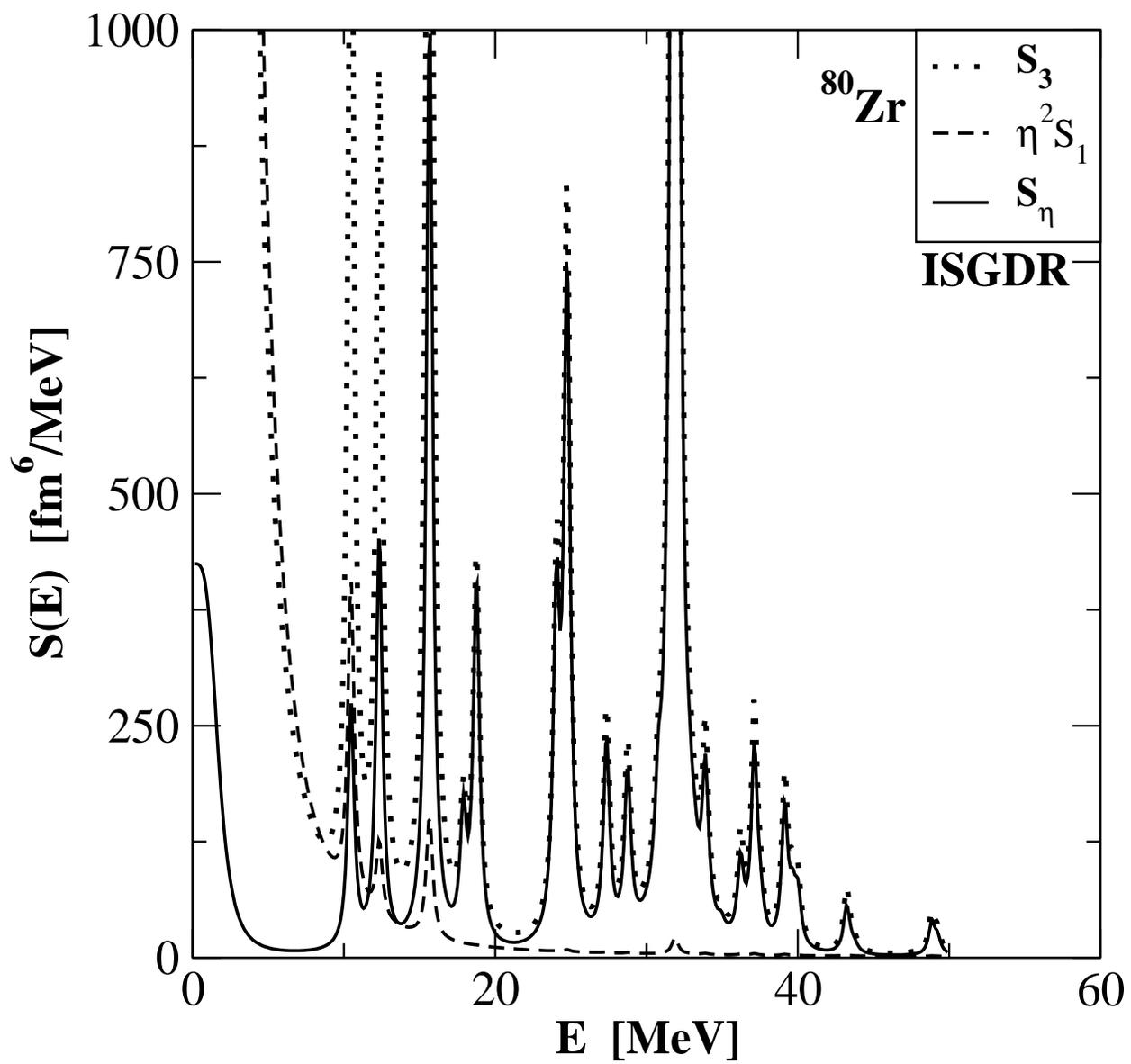
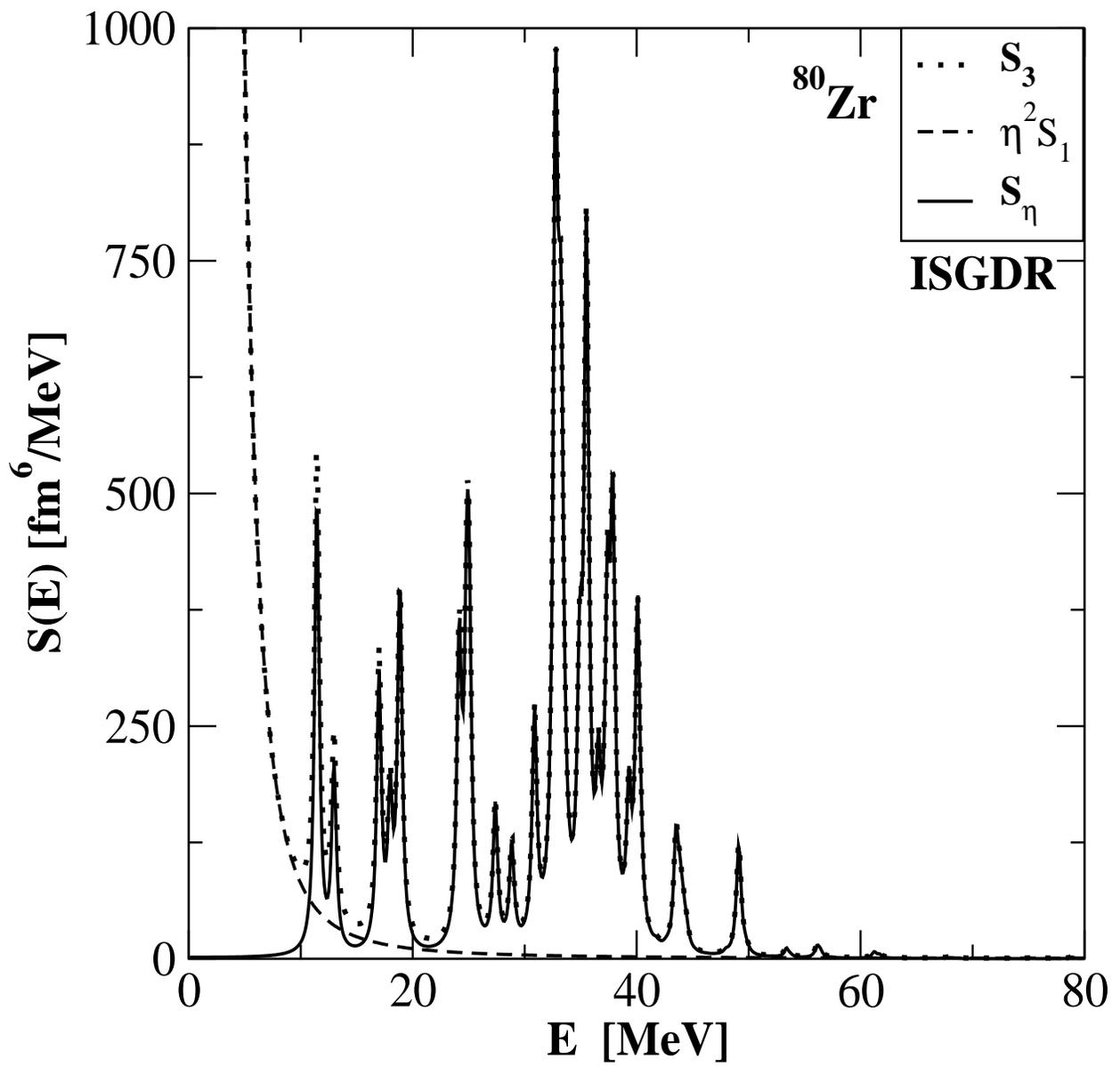
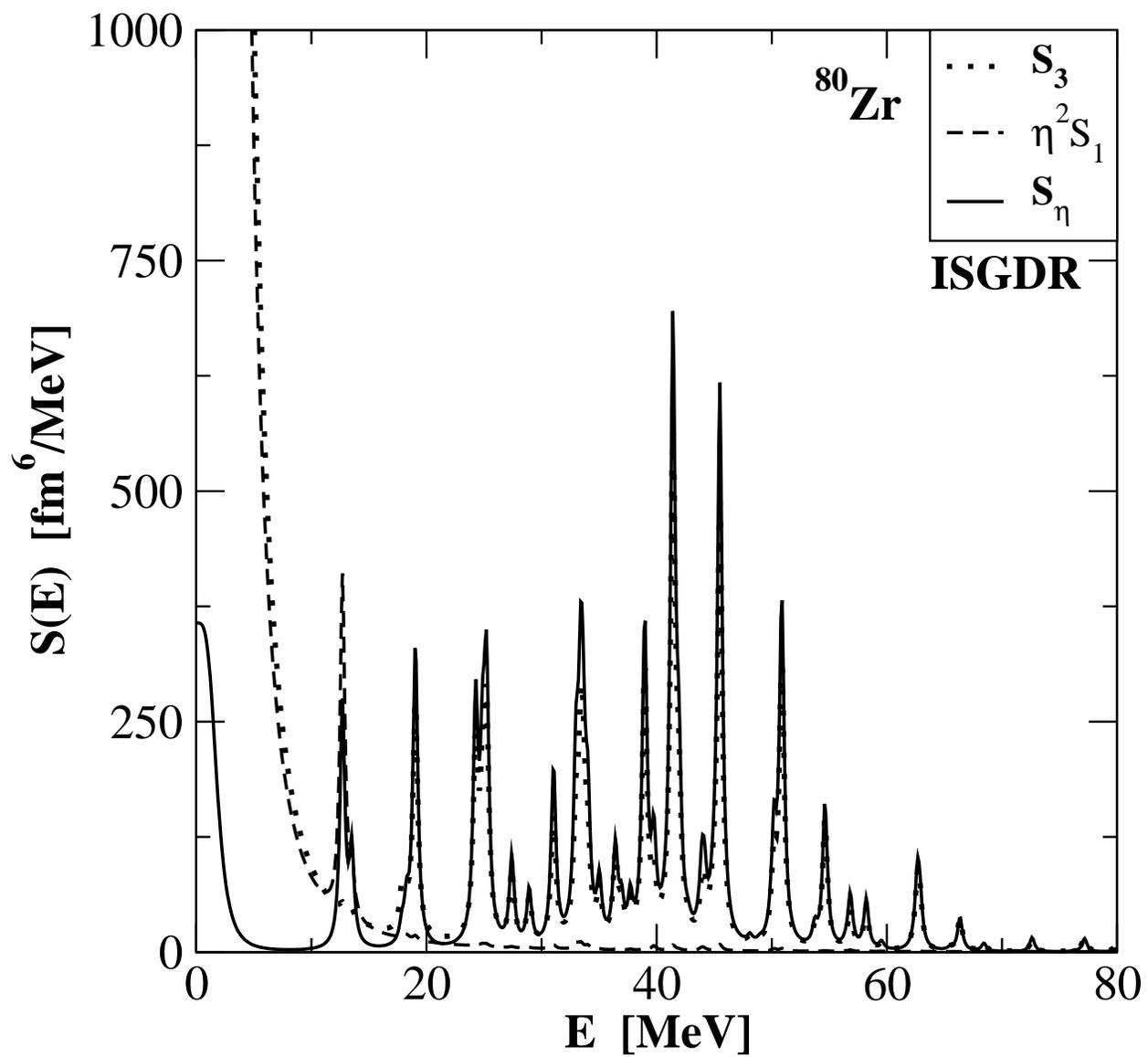
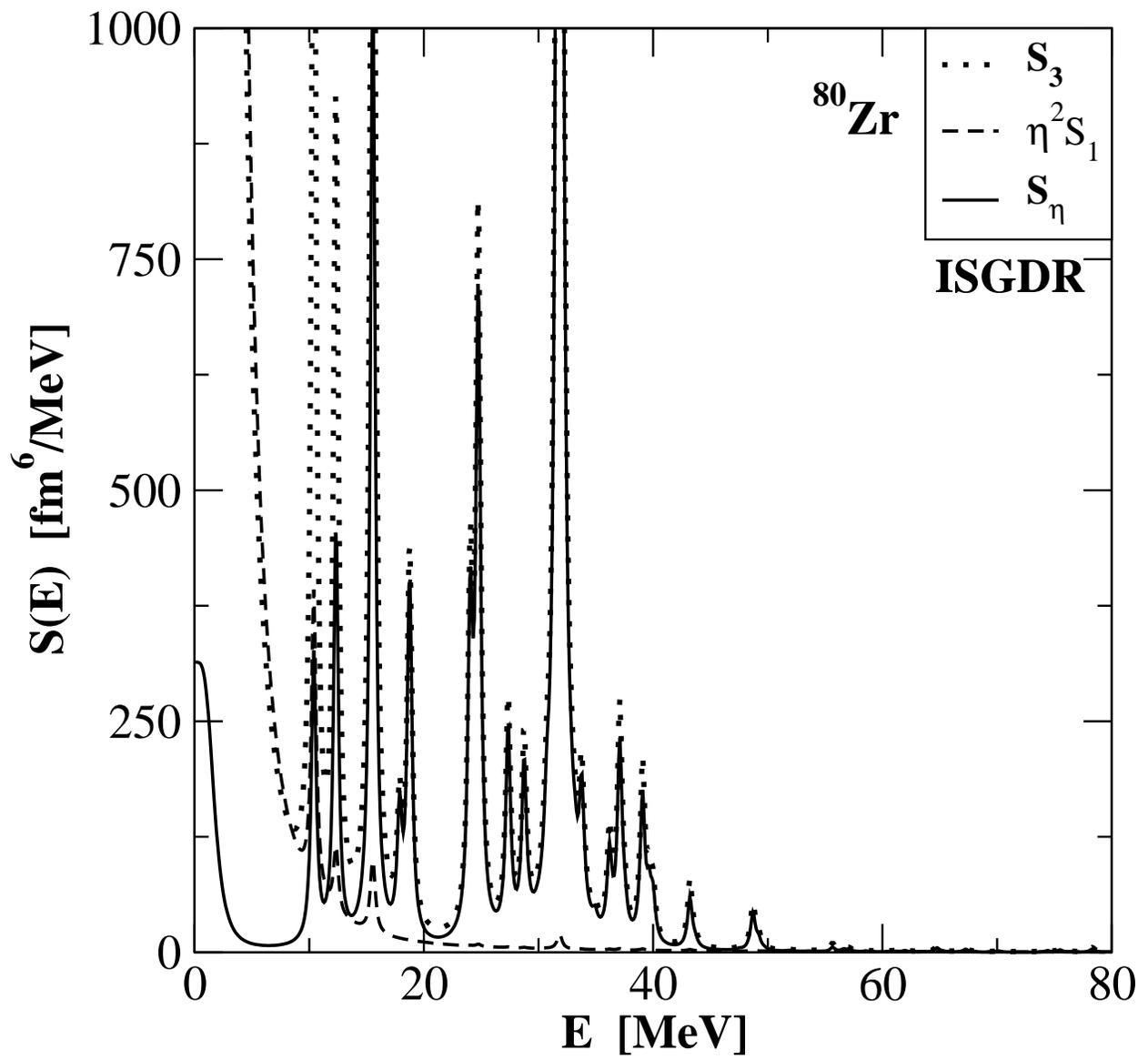
\end{document}